\documentclass[article,12pt,reqno]{amsart}
\usepackage{fullpage}
\usepackage[font=normal,format=plain,labelfont=normal,up]{caption}
\usepackage{hyperref}
\usepackage{amsthm,amsmath}
\usepackage{natbib}
\usepackage[all]{xy}
\usepackage{graphicx}
\usepackage{breqn}
\renewcommand{\baselinestretch}{2}
\allowdisplaybreaks[1]
\makeatletter
                                
\renewcommand{\subsection}{\@startsection{subsection}{2}{\z@}%
                                     {-3.25ex\@plus -1ex \@minus -.2ex}%
                                     {1.5ex \@plus .2ex}%
                                     {\reset@font\normalsize}}

\renewcommand\section{\@startsection {section}{1}{\z@}%
                                   {-3.5ex \@plus -1ex \@minus -.2ex}%
                                   {2.3ex \@plus.2ex}%
                                   {\centering\normalfont\normalsize}}

\makeatother

\renewcommand\title{\normalfont\Large}  

\renewcommand\author{\normalfont\normalsize}   
\everydisplay{
   \abovedisplayskip \baselinestretch\abovedisplayskip%
   \belowdisplayskip \abovedisplayskip%
   \abovedisplayshortskip \baselinestretch\abovedisplayshortskip%
   \belowdisplayshortskip  \baselinestretch\belowdisplayshortskip}
\renewenvironment{abstract}{%
\thispagestyle{empty}%
\centerline{\MakeUppercase{ABSTRACT}}%
}
                                      
\begin{document}

\begin{titlepage}
%\begin{spacing}
\begin{center}
\title{Online Variational Bayes Inference for High-Dimensional Correlated Data}
\end{center}
%\end{spacing}
\vspace{10mm}
\author{
\begin{center}
Sylvie Tchumtchoua, David B. Dunson, and Jeffrey S. Morris
\end{center}
}
\renewcommand{\today}{}
%\thispagestyle{empty}
%\pagebreak
\vspace{5mm}
\begin{abstract}
\vspace{2mm}
\noindent
High-dimensional data with hundreds of thousands of observations are becoming commonplace in many disciplines. The analysis of such data poses many computational challenges, especially when the observations are correlated over time and/or across space. In this paper we propose flexible hierarchical regression models for analyzing such data that accommodate serial and/or spatial correlation. We address the computational challenges involved in fitting these models by adopting an approximate inference framework. We develop an online variational Bayes algorithm that works by incrementally reading the data into memory one portion at a time. The performance of the method is assessed through simulation studies. We applied the methodology to analyze signal intensity in MRI images of subjects with knee osteoarthritis, using data from the Osteoarthritis Initiative. 
\end{abstract}
\thispagestyle{empty}
\end{titlepage}
\pagebreak

\pagenumbering{arabic}

\section{INTRODUCTION}

High-dimensional data arise in a wide range of disciplines, including neuroscience, social and behavioral sciences, bioinformatics, and finance. In this paper we focus on settings where the number of observations per subject is very large relative to the number of subjects and the observations are correlated over time and/or across space. Such data are very popular in medical research, neuroscience and psychology where images consisting of hundreds of thousands of voxels/pixels are collected at several time points on multiple subjects. Conducting statistical analysis on such data poses two key issues. The first issue pertains to the size of the data; statistical methods for analyzing the data all at once are computationally infeasible as they require storing the entire data set into memory, which is impossible with most statistical packages. The second issue is related to accounting for temporal and spatial dependence in the analysis. In image data for example, one expects neighboring and or distant pixels/voxels or regions to have similar neuronal activity or texture information. In addition, sequences of images taken over time are likely to exhibit some temporal correlation. 

Several approaches have been proposed in the literature to overcome these issues. One approach uses a two-step procedure where a linear model is first fitted to each subject's time series at each voxel location separately. In a second stage another regression model is specified with voxel-level regression coefficients as response variables and region of interest (ROI) random effects (\citealp{Bowmanetal08}) or intra ROI regression coefficients and regression coefficients at other stimuli (\citealp{DBK10}) as explanatory variables.  Although this two-step approach on the surface eliminates the sample size problem, it does not adequately model spatial/temporal dependence. The model  in \citet{Bowmanetal08} accounts for between ROI spatial correlation but assumes homogeneous within-region correlation, does not model temporal correlation, and cannot handle very large data sets. On the other hand, the model in \citet{DBK10} is suitable for large data sets but does not model between-region correlation. 

\citet{MC06} proposed  a functional mixed model where the discrete wavelet transform is used to translate the data from the time domain to the frequency domain and all the modeling assumptions and estimation are made in the frequency domain. This work was extended to model image data and use basis functions other than wavelets by \citet{Morrisetal11}. While such an approach makes the computations feasible in moderate to large datasets, by assuming that basis coefficients are independent it restricts the within-function covariance function in a way that is difficult to intuitively grasp and to relate to commonly-used spatial covariance matrices.  

Both the two-step approach and the functional mixed effect model are fitted in the Bayesian framework using Markov chain Monte Carlo (MCMC) techniques that approximate the posterior distribution by repeatedly sampling from the parameters' conditional posterior distributions. Standard MCMC for hierarchical models with longitudinal and/or spatial dependence do not scale well computationally as sample size increases. In addition, assessing convergence of the algorithm can be difficult in complex models. This has motivated various alternative forms of posterior approximation. \citet{RMC09} proposed approximate Bayesian inference for latent Gaussian models by using integrated nested Laplace approximations which is computationally faster than MCMC but their approach does not extend to more flexible models, such as mixtures. 

\citet{WDU11} proposed a fast sequential updating and greedy search algorithm for Dirichlet process mixture models that accommodates very large datasets and does not require reading the entire data into memory but their algorithm relies on the Dirichlet process prediction rule and thus cannot be applied to parametric Bayesian hierarchical models with very large datasets. \citet{Carvalhoetal10} proposed a particle learning approach for mixture models in the state-space framework that builds on the more general framework of \citet{Lopesetal10}. The algorithm approximates the increasing state vector with fixed-dimensional sufficient statistics. \citet{Chopinetal10} showed that the method's performance is poor for large sample sizes unless the number of particles increases exponentially with the number of observations. This makes the algorithm not appropriate for very large datasets.  

Variational Bayes (VB) (\citealp{Jordanetal99}) is another alternative to MCMC that is deterministic and that approximates the posterior distribution with an analytically tractable distribution so that the Kullback-Leiber distance between the complex posterior and its approximation is minimized. The approach typically approximates the posterior with a factorized form for which conjugate priors can be chosen. VB has been used in image analysis and signal processing by several authors including \citet{PKF03}, \citet{OTF10}, \citet{Qietal08}, and \citet{Chengetal05}. The first two studies focus on fMRI time series. Although VB is faster than MCMC for moderate to large datasets, its implementation with very large datasets is computationally expensive as the VB algorithm involves updating observation-specific parameters. Another limitation of VB for very large datasets is that the data is often too large to fit into memory. The key limiting factor for extremely large data is the memory management. Parallel processing can speed up the computations by many factors, but if data cannot be read in, which is the case for many modern applications, then even given hundreds of processors the analysis is a non-starter. 

The objective of this paper is twofold. The first is to develop flexible hierarchical regression models for analyzing very large multiple-subjects data that accommodate spatial and/or temporal correlation. The second is to propose an online VB algorithm that works by reading into memory one portion of the data at a time (for example one image at a time for imaging data), approximating the posterior based on these data and then updating the approximation as additional data are read in.  \citet{HG10} proposed a Bayesian spatio-temporal model for fMRI data where general linear models with an autoregressive error process are fitted to each voxel's time series individually, and a conditional autoregressive prior is specified on the regression  and autoregressive coefficients. To overcome the computational challenges, they used a VB algorithm where the prior distribution at a given iteration depends on the posterior of neighboring coefficients at the previous iteration. Our approach differs from theirs in three respects. First, our approach is a unified framework that models all the voxels at once and is flexible enough to be used with very large data sets with either only spatial or temporal or both spatial and temporal correlation. Second, unlike theirs, our approach is suitable for data collected on several subjects and offers a flexible way to account for heterogeneity among the subjects. Finally, the online aspect of our algorithm refers to reading into memory one part of the data at a time whereas their VB algorithm defines prior distribution sequentially but processes all the coefficients at once. Our online VB algorithm is instead closely related to that of \citet{HBB10} who proposed an online VB algorithm for latent Dirichlet allocation, focusing on the particular class of bag-of-words models for document topics. 

The outline of the paper is as follows. Spatial, temporal, and spatio-temporal semiparametric hierarchical models are developed in the next Section. VB inference is described in Section 3 and an online version of it in Section 4. Simulation examples are given in Section 5, and application to MRI images in Section 6. Section 7 concludes.
 
\section{SEMIPARAMETRIC HIERARCHICAL MODELS}

\subsection{The models}

Let $i=1,...,n$ index subjects, $t=1,...,T$ index time, and $k=1,...,K$ index the spatial units. Let $\mathbf{Y}_{it}$ denote the $K \times 1$ vector of responses for the $i$th subject at time $t$, and $\mathbf{X}_{it}$ be the $K \times g$ matrix of covariates including a column of ones. We specify the model
\begin{equation}\textbf{Y}_{it}=\boldsymbol\eta_{i}\boldsymbol\mu_{it}+\textbf{X}_{it}\boldsymbol\beta_{i}+\boldsymbol\epsilon_{it},\label{eqitj} \end{equation}
where $\boldsymbol\mu_{it}$ is an $m-$dimensional($m<K$) vector of time-varying common factors, $\boldsymbol\eta_{i}$ is a $K \times m$ vector of loadings, and $\boldsymbol\beta_{i}$ is a $g \times 1$ vector of coefficients for subject $i$. $\boldsymbol\epsilon_{it}$ is a vector of error terms assumed independently and identically normally distributed: $\boldsymbol\epsilon_{it} \sim N(\mathbf{0},\sigma^{-2}\mathbf{I})$. 

The first term in the right-hand side of Equation (\ref{eqitj}) specifies a factor model with both latent factors and loadings varying across subjects. In contrast to standard factor analysis where the loadings are typically constant across subjects, we allow the loadings to vary across subjects in order to account for additional heterogeneity among subjects. This specification follows from \citet{AJD02}.  

To model serial correlation, we assume a first-order autoregressive structure for $\boldsymbol\mu_{it}$: 
\begin{align}
\boldsymbol\mu_{it} &\sim N(\boldsymbol\mu_{i,t-1},\theta_{i}^{-1}\textbf{I}),\quad \boldsymbol\mu_{i0}\sim N(\boldsymbol\mu_0,\vartheta\textbf{I}).\label{timco}
\end{align}

To model spatial dependence, we follow the literature on spatial data analysis (see, e.g., \citet{WG09}, \citet{HC03} and \citet{GV03}) and specify a conditional autoregressive model for each column of $\boldsymbol\eta_{i}$. Let $\boldsymbol\eta_{ij}=(\eta_{i1j},\eta_{i2j},...,\eta_{iK})'$ be the loadings on the $j^{th}$ factor. We have
\begin{equation}\boldsymbol\eta_{ij} ~\sim N\left(\mathbf{0},\;\tau^{-1}(I-\rho C)^{-1}\Omega\right),\label{etaij}\nonumber\end{equation}
or, stacking all the columns together,
\begin{equation}\boldsymbol\eta_{i} ~\sim MN_{K \times m}\left(\mathbf{0},\;\mathbf{I}_m,\;\tau^{-1}(I-\rho C)^{-1}\Omega\right),\label{etaij}\end{equation}
where $MN_{K \times m}(.,.,.)$ denotes the matrix normal distribution, $D=(d_{rs})$ denote the proximity matrix, $d_{r+}=\sum_s d_{rs}$, $\Omega=diag\left(\frac{1}{d_{1+}},...,\frac{1}{d_{K+}}\right)$, and $C$ is a $K \times K$ matrix with elements $c_{rs}=\frac{d_{rs}}{d_{r+}}$.

$D=(d_{rs})$ is defined as in \citet{WG09} and \citet{HC03}:
$$d_{rs} = \left\{ \begin{array}{rl} 
			0  &\mbox{if $r=s$,} \\
			\left\|r-s\right\|^{-\phi} &\mbox{otherwise,}\\
\end{array} \right.$$
where $\phi >0$ controls the rate at which the spatial correlation decreases with distance. The value of $\phi$ is chosen so that the loading at a given location only depends on the loadings in a small neighborhood of that location. This results in the matrix $C$ being sparse.

Equations (\ref{eqitj})-(\ref{etaij}) define a model with both spatial and temporal correlation. It closely resembles the spatial dynamic factor model of \citet{LSG08} and the semiparametric dynamic factor model of \citet{Parketal09}, both of which are designed for multivariate time series on a single subject. However our model accommodates multiple subjects and offers a flexible way to accounts for heterogeneity among them in addition to modeling temporal and spatial dependences. Moreover, unlike theirs, our model can be estimated with very large data sets. \citet{Parketal09} applied their model to fMRI data but they overcome the computational challenges by reducing the size of the original images from $64 \times 64 \times 30$ to $32 \times 32 \times 15$. 

The model described by (\ref{eqitj})-(\ref{etaij}) encompasses as special cases models for multivariate time series data with no spatial correlation: \begin{equation}\textbf{Y}_{it}=\boldsymbol\mu_{it}+\textbf{X}_{it}\boldsymbol\beta_{i}+\boldsymbol\epsilon_{it},\label{nospa}\end{equation}
and models for spatial data observed at only few time points used as indicator variables in the design matrix $\mathbf{X}_{it}$: 
\begin{equation}\textbf{Y}_{it}=\boldsymbol\eta_{i}+\textbf{X}_{it}\boldsymbol\beta_{i}+\boldsymbol\epsilon_{it}.\label{notemp}\end{equation}

Finally, a nice property of the factor specification is that temporal and spatial effects are not separable if the number of factor is greater than one (\citealp{LSG08}). A model that uses an additive form $\boldsymbol\mu_{it}+\boldsymbol\eta_{i}$ does not allow spatio-temporal interaction and can be restrictive (\citealp{CH99}). An alternative approach to allowing space-time interaction is to specify a spatial process that evolves over time (\citealp{KDG08}). Although there is a rich recent literature on Gaussian process approximations that scale to reasonably large data sets (refer to \citet{Tokdar07}; \citet{Banerjeeetal08}; \citet{BDT11} among others), such methods are not sufficiently efficient to accommodate our motivating applications. 

\subsection{Prior distributions}

Let $\Theta_i=\left(\boldsymbol\beta_{i}^{'},\theta_i^{'}\right)'$. We flexibly model heterogeneity among subjects by assuming that $\Theta_i$ are drawn from an unknown distribution which has the Dirichlet process prior:
\begin{align*}
\Theta_i &\sim G,\quad G = \sum_{r=1}^\infty \pi_r\delta_{\Theta_r^{*}},\quad \pi_r =v_r\prod_{l<r}(1-v_l),\quad \Theta_r^{*}=\left(\boldsymbol\beta_r^{*'}, \theta_r^{*'}\right)',\\
v_r &\sim Beta(1,\alpha),\quad \alpha \sim Ga(a_{\alpha},b_{\alpha}), \quad \boldsymbol\beta_r^{*} \sim N(\boldsymbol\beta_{0r},\Sigma_{0r}),\; \theta_r^{*} \sim Ga(a_{\theta},b_{\theta}). 
\end{align*}

For the other parameters we use $\sigma^2 \sim Ga(a_{\sigma},b_{\sigma})$, $\tau \sim Ga(a_{\tau},b_{\tau})$. In order to simplify computations, we follow \citet{GV03} in discretizing $\rho$ and assume it takes values $\rho_{l}=\frac{l}{M}, \; l=0,1,...,M-1,M-\epsilon$ with equal probability: $\phi_{l}=Pr(\rho=\rho_{l})=\frac{1}{M+1}$.

One could use MCMC techniques (details in Appendix~\ref{MCMC}) to estimate the parameters ${\beta_{i}}$, ${\theta_{i}}$, ${\boldsymbol\mu_{it}}$, $\boldsymbol\eta_{ij}$, $\tau$, $\rho$, and $\sigma^2$ but this is not practical for large datasets. In the next Section we derive variational Bayes inference for the models. 

\section{VARIATIONAL BAYES INFERENCE}

\citet{BJ06} proposed a variational Bayes inference algorithm for Dirichlet process mixtures which was implemented by \citet{Qietal08} in the context of multi-task compressive sensing. We adapt their algorithm to the spatio-temporal setup.

Define the allocation variables $z_{i}$ so that $z_{i}=r$ if $\Theta_{i}=\Theta_r^{*}$. 

The variational distribution $q(V,\boldsymbol\Theta^{*},Z,\boldsymbol\eta,\boldsymbol\mu,\rho,\tau,\sigma^2,\alpha)$ is defined as 
\begin{eqnarray*}q(V,\boldsymbol\Theta^{*},Z,\boldsymbol\eta,\boldsymbol\mu,\rho,\tau,\sigma^2,\alpha)&=&q(\sigma^2) q(\alpha)q(\rho)q(\tau)\left(\prod_{r=1}^R q(\theta_r^{*})q(\boldsymbol\beta_r^{*})q(v_r)\right)\left(\prod_{i=1}^n q(z_{i})\right)\nonumber \\
&\times &\left(\prod_{i=1}^n \prod_{t=1}^T q(\mu_{it})\right)\left(\prod_{i=1}^nq(\boldsymbol\eta_{i})\right),\label{varia} \end{eqnarray*}
where 
$q(\sigma^2)= Ga(\sigma^2;\tilde{a}_{\sigma},\tilde{b}_{\sigma})$, $q(\alpha)= Ga(\alpha;\tilde{a}_{\alpha},\tilde{b}_{\alpha})$,  $q(\rho)= Mult(\rho;\pi_{1},..,\pi_{M+1})$, $q(\tau)= Ga(\tau;\tilde{a}_{\tau},\tilde{b}_{\tau})$, 
$q(\theta_r^{*})= Ga(\theta_r^{*};\tilde{a}_{\theta_r},\tilde{b}_{\theta_r})$, $q(\boldsymbol\beta_r^{*})= N(\boldsymbol\beta_r^{*};\tilde{\boldsymbol\beta}_{0r},\tilde{\Sigma}_{0r})$,  $q(v_r)= Be(v_r;\gamma_{r1},\gamma_{r2})$ with $q(v_R=1)=1$,  $q(z_{i})= Mult(z_{i};\kappa_{i1},..,\kappa_{iR})$, $q(\boldsymbol\mu_{it})= N(\boldsymbol\mu_{it};\boldsymbol\lambda_{it,1},\boldsymbol\lambda_{it,2})$, $q(\boldsymbol\eta_{i})= MN_{K \times m}(\boldsymbol\eta_{i};\boldsymbol\xi_{i},\mathbf{I}_m,\boldsymbol\Psi_{i})$, $\boldsymbol\xi_{i}=\left[\boldsymbol\xi_{i1},...,\boldsymbol\xi_{im}\right]$, $\boldsymbol\Psi_{i}=diag(\psi_{il})$, $l=1,...,K$.

The variational objective function and update equations are given in Appendix~\ref{objectiveVB} and Appendix~\ref{updateequationVB}, respectively. The variational Bayes algorithm iterates the update equations until the objective function converges.

The implementation of the variational Bayes algorithm requires reading into memory all the data at once and involves updating observation-specific parameters. With very large datasets, one may not be able to read in all the data at once. We next describe an online version of the algorithm that works by incrementally reading the data into memory one portion at a time.

\section{ONLINE VARIATIONAL BAYES INFERENCE}

Few approaches for online learning of Bayesian mixture models have been proposed in the literature. \citet{Sato01} proposed an online variational Bayes algorithm for mixture models where the amount of data increases over time and  a time-dependent discount factor is used to  decay the terms of the objective function that correspond to old data. A similar approach was used by \citet{HV03} in online variational Bayesian learning using linear independent component analysis and more recently by \citet{HBB10} in online learning for latent Dirichlet allocation (LDA).

\citet{Fear04} proposed a particle filter algorithm for Dirichlet process mixture models where the number of clusters increases with each new data point observed but a fixed number of particles are used to approximate the posterior distribution. \citet{GWP08} proposed a ``memory bounded variational Dirichlet process" for online category discovery where data are processed in small batches. In each batch, a standard VB is first used to determine the clustering estimates; then in a compression phase partitions are repeatedly split and the data points assigned to the same cluster are summarized by the cluster sufficient statistics and the original data are discarded. 

We use the discounting approach of \citet{Sato01}. The data matrix $(\mathbf{Y},\mathbf{X})$ is viewed as a set of $n$ samples $\left\{\mathbf{Y}_{it},\mathbf{X}_{it}\right\}_{t=1}^T$, $i=1,...,n$.  

Let $(\mathbf{Y}^s,\mathbf{X}^s)=\left\{\mathbf{Y}_{it},\mathbf{X}_{it},\; i=1,...,s;\;t=1,...,T\right\}$  represent the data set up to and including subject $s<n$. The online variational algorithm processes the data one subject at a time. Let $q_s(\sigma^2)$, $q_s(\alpha)$,  $q_s(\rho)$, $q_s(\tau)$, $q_s(\theta_r^{*})$, $q_s(\boldsymbol\beta_r^{*})$, and $q_s(v_r)$  denote the estimates based on the observed data $(\mathbf{Y}^s,\mathbf{X}^s)$, and $A=(a_{\alpha},b_{\alpha},a_{\theta},b_{\theta},a_{\tau},b_{\tau},\boldsymbol\mu_0,\vartheta)$ the known hyper-parameters. 

The discounted objective function based on $s$ observations takes the form 
\begin{eqnarray*}
\ell^s(\mathbf{Y}^s|\mathbf{X}^s,A) &=& E_q\left[log\;p(\sigma^2)-log\;q(\sigma^2)\right]+E_q\left[log\;p(\alpha)-log\;q(\alpha)\right]\nonumber \\
&+& E_q\left[log\;p(\rho)-log\;q(\rho)\right]+E_q\left[log\;p(\tau)-log\;q(\tau)\right]\nonumber \\
&+& E_q\left[log\;p(V|\alpha)-log\;q(V)\right] +E_q\left[log\;p(\theta^{*})-log\;q(\theta^{*})\right] +E_q\left[log\;p(\boldsymbol\beta^{*})-log\;q(\boldsymbol\beta^{*})\right]\nonumber \\
&+&
\sum_{i=1}^s d(i,s)\left\{\sum_{t=1}^T E_q\left[log\;p(z_{it}|V)-log\;q(z_{it})\right]\right\} \nonumber \\
&+& \sum_{i=1}^s d(i,s)\left\{\sum_{t=1}^T E_q\left[log\;p(\boldsymbol\mu_{it}|\boldsymbol\mu_{i,t-1},z_{it})-log\;q(\boldsymbol\mu_{it})\right]\right\}\nonumber \\
&+& \sum_{i=1}^s d(i,s)\left\{E_q\left[log\;p(\boldsymbol\eta_{i})-log\;q(\boldsymbol\eta_{i})\right]\right\}\nonumber \\
&+& \sum_{i=1}^s d(i,s) \left\{\sum_{t=1}^T E_q\left[log\;p(\mathbf{Y}_{it}|\boldsymbol\mu_{it},\;\boldsymbol\eta_{i},\sigma^2,A)\right]\right\}. \label{llb}
\end{eqnarray*}

where $0\leq d(i,s)\leq 1$ are discount factors defined as
\begin{equation}d(i,s)=\prod_{l=i+1}^s(1-h(l)),\;d(s,s)=1. \label{disc} \end{equation}

$h(l)$ is a known function satisfying $0\leq h(l)\leq 1$. 

The online algorithm proceeds by repeating the following steps as new data arrive:
\begin{enumerate}
\item when a new sample $\left\{\mathbf{Y}_{s+1,t},\mathbf{X}_{s+1,t},\; t=1,...,T\right\}$ arrives, $\ell^{s+1}(\mathbf{Y}^{s+1}|\mathbf{X}^{s+1},A)$ is maximixed with respect to $q_{s+1}(z_{s+1})$, $q_{s+1}(\boldsymbol\mu_{s+1})$ and $q_{s+1}(\boldsymbol\eta_{s+1})$ while $q(\sigma^2)$, $q(\alpha)$,  $q(\rho)$, $q(\tau)$, $q(\theta_r^{*})$, $q(\boldsymbol\beta_r^{*})$, and $q(v_r)$ are set to $q_s(\sigma^2)$, $q_s(\alpha)$,  $q_s(\rho)$, $q_s(\tau)$, $q_s(\theta_r^{*})$, $q_s(\boldsymbol\beta_r^{*})$, and $q_s(v_r)$, respectively. 
\item The discounted objective function is next maximized with respect to $q(\sigma^2)$, $q(\alpha)$,  $q(\rho)$, $q(\tau)$, $q(\theta_r^{*})$, $q(\boldsymbol\beta_r^{*})$, and $q(v_r)$,  while $q(z_{i})$, $q(\boldsymbol\mu_{it})$ and $q(\boldsymbol\eta_{i})$, $i=1,...,s+1$ are fixed. 
\end{enumerate}

The following update equations are obtained:
\begin{enumerate}
\item Subject-specific parameters
\begin{itemize}
\item $\boldsymbol\lambda_{s+1,0,2}=\left(\vartheta^{-1}+\sum_{r} \kappa_{s+1,r} \frac{\tilde{a}_{\theta_r}^s}{\tilde{b}_{\theta_r}^s}\right)^{-1},$ \\
$\boldsymbol\lambda_{s+1,0,1}=\boldsymbol\lambda_{s+1,0,2}\left(\boldsymbol\lambda_{s+1,1,1} \sum_{r} \kappa_{s+1,r} \frac{\tilde{a}_{\theta_r}^s}{\tilde{b}_{\theta_r}^s}+\boldsymbol\mu_0\vartheta^{-1}\right),$\\
$\boldsymbol\lambda_{s+1,t,2}=\left((tr(\boldsymbol\Psi_{s+1})+\boldsymbol\xi_{s+1}'\boldsymbol\xi_{s+1})\frac{\tilde{a}_{\sigma}^s}{\tilde{b}_{\sigma}^s}+\sum_{r} \kappa_{s+1,r} \frac{\tilde{a}_{\theta_r}^s}{\tilde{b}_{\theta_r}^s}\right)^{-1},$   \\
$\boldsymbol\lambda_{s+1,t,1}=\boldsymbol\lambda_{s+1,t,2}\sum_{r}\ \kappa_{s+1,r}\left(\boldsymbol\xi_{s+1}'(\mathbf{Y}_{s+1,t}-\mathbf{X}_{s+1,t}\boldsymbol\beta_{0r}^s)\frac{\tilde{a}_{\sigma}^s}{\tilde{b}_{\sigma}^s}+\boldsymbol\lambda_{s+1,t-1,1}\frac{\tilde{a}_{\theta_r}^s}{\tilde{b}_{\theta_r}^s}\right).$ 
\item $\kappa_{s+1,r}\propto exp(w_{s+1,r})$,\\ $w_{s+1,r}=-\frac{\tilde{a}_{\sigma}^s}{2\tilde{b}_{\sigma}^s}\sum_t\left\{\tilde{\mathbf{Y}}_{s+1,t}'\tilde{\mathbf{Y}}_{s+1,t}+(tr(\boldsymbol\Psi_{s+1})+\boldsymbol\xi_{s+1}'\boldsymbol\xi_{s+1})tr(\boldsymbol\lambda_{s+1,t,2})\right\}$ \\$-$ $\frac{\tilde{a}_{\sigma}^s}{2\tilde{b}_{\sigma}^s}\sum_t\left\{\lambda_{s+1,t,1}'tr(\boldsymbol\Psi_{s+1})\boldsymbol\lambda_{s+1,t,1}+ tr\left(\mathbf{X}_{s+1,t}'\mathbf{X}_{s+1,t}\tilde{\Sigma}_{0r}^s\right)\right\}$ \\$-$ $0.5\frac{\tilde{a}_{\theta_r}^s}{\tilde{b}_{\theta_r}^s}\sum_t\left\{(\boldsymbol\lambda_{s+1,t,1}-\boldsymbol\lambda_{s+1,t-1,1})'(\boldsymbol\lambda_{s+1,t,1}-\boldsymbol\lambda_{s+1,t-1,1})+tr(\boldsymbol\lambda_{s+1,t,2}+\boldsymbol\lambda_{s+1,t-1,2})\right\}$ \\$+$ $\psi(\gamma_{r1}^s)-\psi(\gamma_{r1}^s+\gamma_{r2}^s)+\sum_{l=1}^{r-1} \left\{\psi(\gamma_{l2}^s)-\psi(\gamma_{l1}^s+\gamma_{l2}^s)\right\}$ $+$ $\frac{KT}{2}(\psi(\tilde{a}_{\sigma}^s)-log(\tilde{b}_{\sigma}^s)),$ \\
where $\tilde{\mathbf{Y}}_{s+1,t}=\mathbf{Y}_{s+1,t}-\boldsymbol\xi_{s+1}\boldsymbol\lambda_{s+1,t,1}-\mathbf{X}_{s+1,t}\tilde{\boldsymbol\beta}_{0r}^s$ and $\psi(.)$ is the digamma function.
\item $\psi_{s+1,k}=\left(T\frac{\tilde{a}_{\sigma}^s}{\tilde{b}_{\sigma}^s}+\frac{\tilde{a}_{\tau}^s}{\tilde{b}_{\tau}^s}\Omega_{kk}^{-1}\right)^{-1}$, $k=1,...,K$,\\
      $\xi_{ijk}=\psi_{ik}\frac{\tilde{a}_{\sigma}^s}{\tilde{b}_{\sigma}^s}\sum_r\kappa_{s+1,r}\sum_t \lambda_{s+1,kt,1}(\mathbf{Y}_{s+1,kt}-\mathbf{X}_{s+1,kt}\tilde{\boldsymbol\beta}_{0r}^s)$, $j=1,...,m$, $k=1,...,K$
\end{itemize}
\item Global parameters 
\begin{itemize}     
\item $\tilde{a}_{\sigma}^{(s+1)}=(1-h(s+1))\tilde{a}_{\sigma}^{s}+h(s+1)\left(a_{\sigma}+\frac{KT}{2h(s+1)}\right),$ 
\begin{dmath*}
      \tilde{b}_{\sigma}^{(s+1)}=(1-h(s+1))\tilde{b}_{\sigma}^{s}+h(s+1)\left(
b_{\sigma}+\frac{1}{2h(s+1)}\sum_{t,r}\kappa_{s+1,r}\left\{\tilde{\mathbf{Y}}_{s+1,t}'\tilde{\mathbf{Y}}_{s+1,t} 
+(tr(\boldsymbol\Psi_{s+1})+\boldsymbol\xi_{s+1}'\boldsymbol\xi_{s+1})tr(\boldsymbol\lambda_{s+1,t,2})+\lambda_{s+1,t,1}'tr(\boldsymbol\Psi_{s+1})\boldsymbol\lambda_{s+1,t,1}+tr\left(\mathbf{X}_{s+1,t}'\mathbf{X}_{s+1,t}\tilde{\Sigma}_{0r}^{s+1}\right)\right\}\right),
\end{dmath*}
where $\tilde{\mathbf{Y}}_{s+1,t}=\mathbf{Y}_{s+1,t}-\boldsymbol\xi_{s+1,}\boldsymbol\lambda_{s+1,t,1}-\mathbf{X}_{s+1,t}\tilde{\boldsymbol\beta}_{0r}^{s+1}.$
\item $\tilde{a}_{\alpha}^{(s+1)} = c+R-1,$\\
      $\tilde{b}_{\alpha}^{(s+1)}=b_{\alpha}-\sum_{r=1}^{R-1} (\psi(\gamma_{r2}^{(s+1)})-\psi(\gamma_{r1}^{(s+1)}+\gamma_{r2}^{(s+1)}))$
\item $\tilde{a}_{\theta_r}^{(s+1)} = (1-h(s+1))\tilde{a}_{\theta_r}^{s} +h(s+1)\left(a_{\theta}+\frac{T}{2h(s+1)}\kappa_{s+1,r}\right),$
\begin{dmath*}
      \tilde{b}_{\theta_r}^{(s+1)} = (1-h(s+1))\tilde{b}_{\theta_r}^{s}  
+h(s+1)\left(b_{\theta}+\frac{1}{2h(s+1)}\sum_{t=1}^T\kappa_{s+1,r}\left\{(\boldsymbol\lambda_{s+1,t,1}-\boldsymbol\lambda_{s+1,,t-1,1})'(\boldsymbol\lambda_{s+1,t,1}-\boldsymbol\lambda_{s+1,t-1,1})+tr(\boldsymbol\lambda_{s+1,t,2}+\boldsymbol\lambda_{s+1,,t-1,2})\right\}\right).
\end{dmath*}
\item $\left(\tilde{\Sigma}_{0r}^{(s+1)}\right)^{-1}=(1-h(s+1))\left(\tilde{\Sigma}_{0r}^{s}\right)^{-1}$ $+$ $h(s+1)\left(\Sigma_0^{-1}+\frac{\tilde{a}_{\sigma}^{(s+1)}}{h(s+1)\tilde{b}_{\sigma}^{(s+1)}}\sum_t\kappa_{s+1,r}\mathbf{X}_{s+1,t}'\mathbf{X}_{s+1,t}\right)$,\\
  $\tilde{\boldsymbol\beta}_{0r}^{(s+1)}=(1-h(s+1))\tilde{\Sigma}_{0r}^{(s+1)}\left(\tilde{\Sigma}_{0r}^{s}\right)^{-1}\tilde{\boldsymbol\beta}_{0r}^{s}$\\
		$+$ $h(s+1)\tilde{\Sigma}_{0r}^{(s+1)}\left(\boldsymbol\beta_{0}'\Sigma_0^{-1}+\frac{\tilde{a}_{\sigma}^{(s+1)}}{h(s+1)\tilde{b}_{\sigma}^{(s+1)}}\sum_t\kappa_{s+1,r}(\mathbf{Y}_{s+1,t}-\boldsymbol\xi_{s+1}\boldsymbol\lambda_{s+1,t,1})'\mathbf{X}_{s+1,t}\right).$    
\item $\gamma_{r1}^{(s+1)} = (1-h(s+1))\gamma_{r1}^{s} +h(s+1)\left(1+\frac{1}{h(s+1)}\kappa_{s+1,r}\right)$ \\
      $\gamma_{r2}^{(s+1)} = (1-h(s+1))\gamma_{r2}^{s} +h(s+1)\left(\frac{\tilde{a}_{\alpha}^{(s+1)}}{\tilde{b}_{\alpha}^{(s+1)}}+\frac{1}{h(s+1)} \sum_{l=r+1}^R \kappa_{s+1,l}\right)$
\item $\tilde{a}_{\tau}^{s+1}=(1-h(s+1))\tilde{a}_{\tau}^s+h(s+1)(a_{\tau}+\frac{1}{2h(s+1)}K_j)$,
      $\tilde{b}_{\tau}^{s+1}=(1-h(s+1))\tilde{b}_{\tau}^s+h(s+1)(b_{\tau}+\frac{1}{2h(s+1)}\sum_l \pi_{l}\left\{m\:tr(\Omega^{-1}\boldsymbol\Psi_{s+1})+\sum_{k=1}^m\boldsymbol\xi_{s+1,k}'\Omega^{-1}(I-\rho_l C)\boldsymbol\xi_{s+1,k}\right\}.$ 
\item $\tilde{\pi}_{l}^{(s+1)} \propto exp(\varpi_{l}^{(s+1)})$,
\begin{dmath*}
      \varpi_{l}^{(s+1)}=(1-h(s+1))log(\varpi_{l}^s)+0.5m\:n\:K\:(\psi(\tilde{a}_{\tau}^{(s+1)})-log(\tilde{b}_{\tau}^{(s+1)})) -0.5m\: (log(\left|\Omega\right|)- log\left(\left|(I-\rho_{l} C) \right|\right))  
      -\frac{a_{\tau}^{(s+1)}}{2b_{\tau}^{(s+1)}}\left\{m\:tr(\Omega^{-1}\boldsymbol\Psi_{s+1})+ \sum_{k=1}^m\boldsymbol\xi_{s+1,k}'\Omega^{-1}(I-\rho_{l} C)\boldsymbol\xi_{s+1,k}\right\}.
\end{dmath*}
Owing to the sparsity of the matrix $C$, $log\left(\left|(I-\rho_{l} C) \right|\right)$ is rapidly computed, even for very large values of $K$, using the methods described by \citet{BPa99}.        
\end{itemize}
\end{enumerate}Given a set of starting values, each step is iterated until the changes in the estimates at two consecutive iterations are small. 

Notice that $\tilde{a}^{(s+1)}$, $\tilde{b}^{(s+1)}$, $\gamma_{r1}^{(s+1)}$, $\gamma_{r2}^{(s+1)}$, $\tau_{r1}^{(s+1)}$, and $\tau_{r2}^{(s+1)}$ are weighted averages of their previous values and the estimates based solely on the current data. By specifying $h(s+1)=\frac{1}{s+1}$, these estimates are based on the current data repeated $s+1$ times. 

A key target for inference in model (\ref{eqitj})-(\ref{etaij}) is the predictive distribution of $\boldsymbol\beta_{n+1}$ for an additional individual, which is obtained as:
\begin{equation}p(\boldsymbol\beta_{n+1}|a_{\alpha},b_{\alpha},\boldsymbol\beta_{0},\Sigma_{0},\left\{\boldsymbol\beta_{i}\right\}_{i=1}^n,\textbf{X},\textbf{Y})\approx \sum_{r=1}^R\left(\frac{\gamma_{1r}}{\gamma_{1r}+\gamma_{2r}}\prod_{l<r}\frac{\gamma_{1l}}{\gamma_{1l}+\gamma_{2l}}\right) N(\tilde{\boldsymbol\beta}_{0r},\tilde{\Sigma}_{0r}). \nonumber \end{equation}

\section{SIMULATED EXAMPLES}

\subsection{Simulated example 1: time series, no spatial dependence}

We generated $n=10,000$ time series, each of length $T=50$ as follows: 
\begin{align*}
Y_{it} &= \mu_{it} +\epsilon_{it},\; i=1,...,n,\; t=1,...,T\\
\mu_{it} &\sim N(\mu_{i,t-1},\theta_{i}),\;\mu_{i0} \sim N(0,1)\\
\theta_{i}&=\left\{ \begin{array}{rl} 
\theta_1^{*} \sim Ga(2,1/3) &\mbox{w.p. $1/2$}\\
\theta_2^{*} \sim Ga(4,1/5) &\mbox{w.p. $1/2$}
\end{array} \right.\\
\epsilon_{it} &\sim N_K(0,1/7)
\end{align*}
We applied the online algorithm described in the previous Section, reading in the data one time series at a time. The hyperparameters $a_{\sigma}$, $b_{\sigma}$, $a_{\alpha}$, $b_{\alpha}$, $a_{\theta}$, and $b_{\theta}$ were specified as: $a_{\sigma}=b_{\sigma}=a_{\alpha}=b_{\alpha}=1$ and $a_{\theta}=b_{\theta}=10^{-4}$, and the truncation level of the stick-breaking process was fixed at $R=20$. We used $h(l)=1/l$. The variational distributions were initialized by setting their parameters to the hyperparameters of the corresponding prior distributions, and the algorithm was iterated until the changes in the estimates at two consecutive iterations was less that $1e^{-6}$. It took 2.53 hours on a Windows operated laptop with 2.27 Ghz and 4 GB RAM using Matlab to analyze the entire data. 

\begin{figure} 
%\vspace{6pc}
\centering
\includegraphics[scale=0.5]{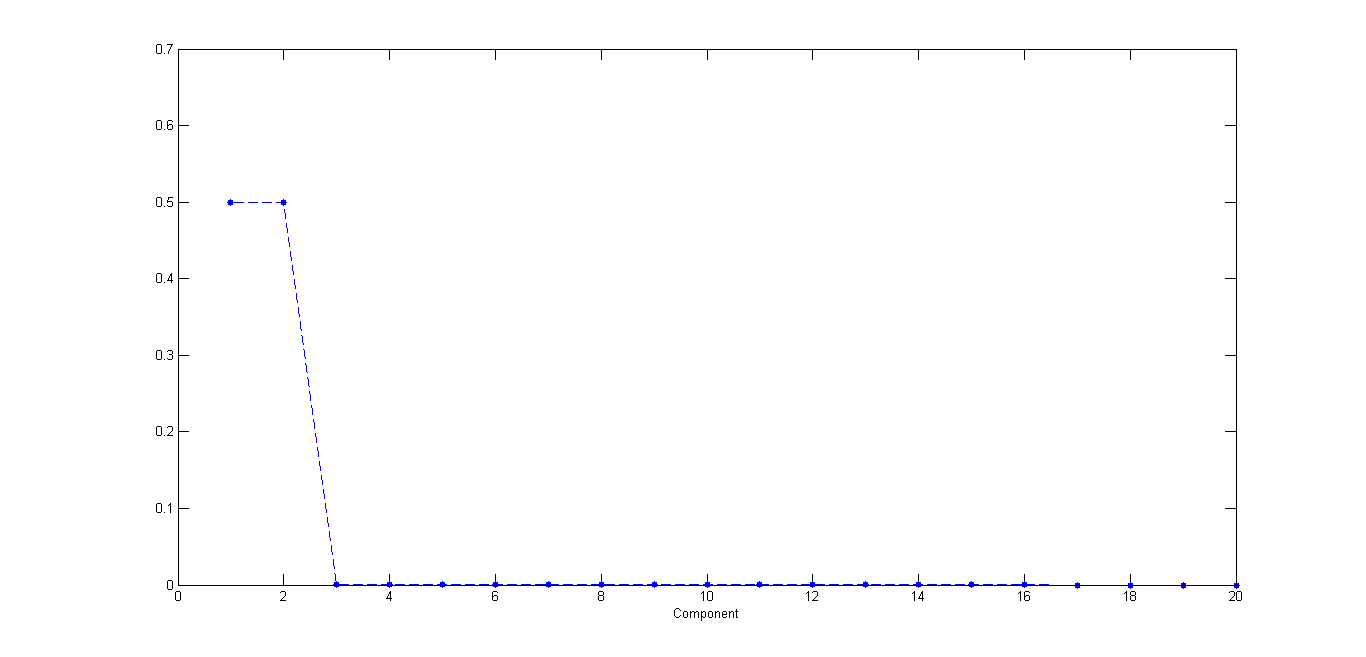}
\caption[]{Estimated mixing proportions $\pi_r$, $r=1,...,20$.}
\label{pi}
\end{figure}

\begin{figure} 
\centering
\includegraphics[scale=0.5]{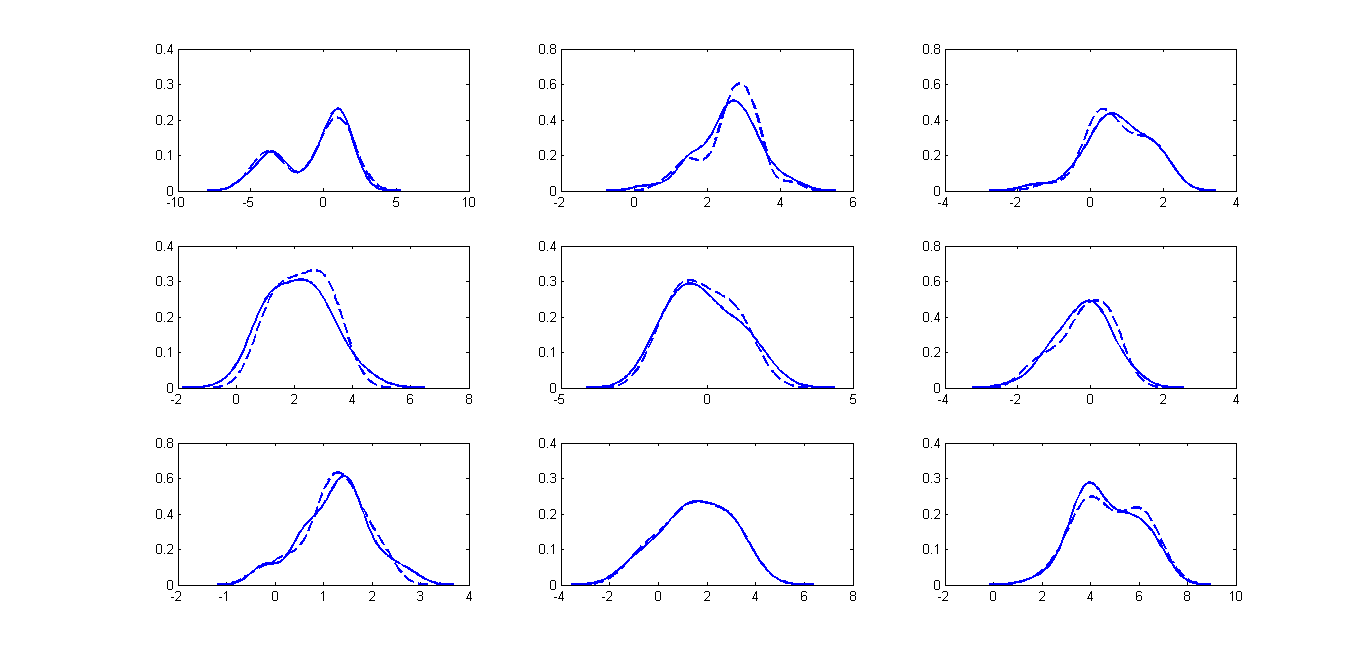}
\caption[]{Simulated example 1. Kernel smoothed density estimates of $\mu_{it}$ for the last 9 observations. The dashed lines are the distributions of the true values and the solid lines are the distributions of the online VB estimates.}
\label{predicted9}
\end{figure}

Figure~\ref{pi} plots the expected mixing proportions $E_q\left[\pi_r(V)\right]=\frac{\gamma_{1r}}{\gamma_{1r}+\gamma_{2r}}\prod_{l<r}\frac{\gamma_{1l}}{\gamma_{1l}+\gamma_{2l}}$. The model correctly identifies two components with equal weight.   
In Figure~\ref{predicted9} are displayed the distributions of the parameters $\mu_{it}$ for each of the last 9 time series. The variational distributions approximate the true distributions quite well. 

\subsection{Simulated example 2: spatial model, no temporal dependence} 

We generated data from the spatial hierarchical model 
\begin{align*}
\textbf{Y}_{it}&=\boldsymbol\eta_{i}+\textbf{X}_{it}\boldsymbol\beta_{i}+\boldsymbol\epsilon_{it},\;\boldsymbol\eta_{i} \sim N_K(\mathbf{0},\mathbf{I}),\; \boldsymbol\epsilon_{it} \sim N_K(\mathbf{0},\mathbf{I}),\;i=1,...,n,\; t=1,...,T\\
\boldsymbol\beta_{i}&=\left\{ \begin{array}{rl} 
\boldsymbol\beta_1^{*} \sim N(\boldsymbol\beta_{01},\Sigma) &\mbox{w.p. $1/2$}\\
\boldsymbol\beta_2^{*} \sim N(\boldsymbol\beta_{02},\Sigma) &\mbox{w.p. $1/2$}
\end{array} \right.\\
\end{align*}
where $K=65,536$, $n=400$, $T=5$, $\boldsymbol\beta_{01}=(1.5, 1.5, 1, 2, 2)'$, $\boldsymbol\beta_{02}=(-1.5, -1.5, -1, -2, -2)'$, $\Sigma^{-1} \sim  W(I,10)$, $\textbf{X}_{it}$ is a $K \times T$ matrix whose columns are indicator variables for time.

The online algorithm was implemented reading in the data one subject at a time. Hyperparameter values and the stopping criterion were chosen as in Simulated example 1. The run time was 2.25 hours. 

\begin{figure} 
\centering
\includegraphics[scale=0.5]{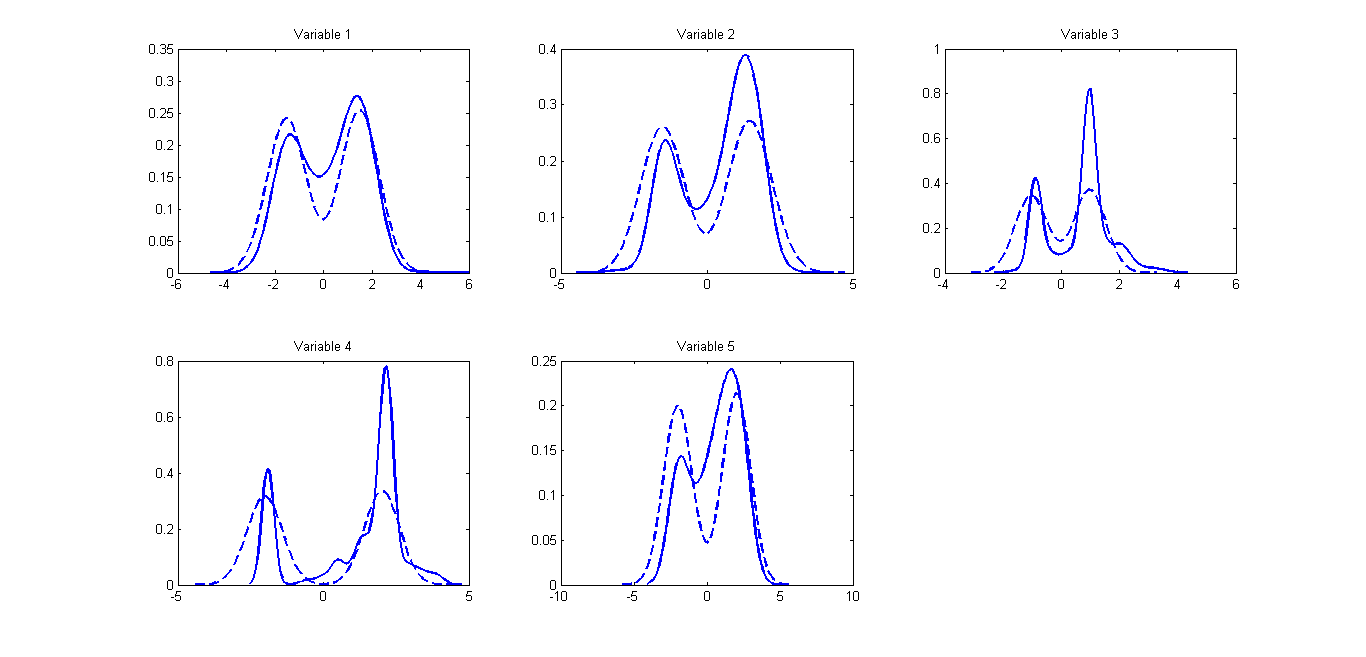}
\caption[]{Simulated example 2. Online VB approximate (solid lines) predictive densities of components of $\boldsymbol\beta_{n+1}$ and the associated true densities (dashed lines) for $K=65,536$ and $n=400$.}
\label{predicted400}
\end{figure}

The online VB estimates of $\sigma^2$ is 1.121 with a 95\% credible interval of [1.119, 1.123]. Those of $\rho$ and $\tau$ are 0.586([0.115, 0.877]) and 0.0421([0.0420, 0.0423]), respectively. 
Figure~\ref{predicted400} shows the online VB approximations (solid lines) to the predictive densities of components of $\boldsymbol\beta_{n+1}$ and the associated true densities (dashed lines). Online VB is able to recover the shape of the true distributions and correctly estimate the location of the two prominent modes.  

\begin{figure} 
\centering
\includegraphics[scale=0.5]{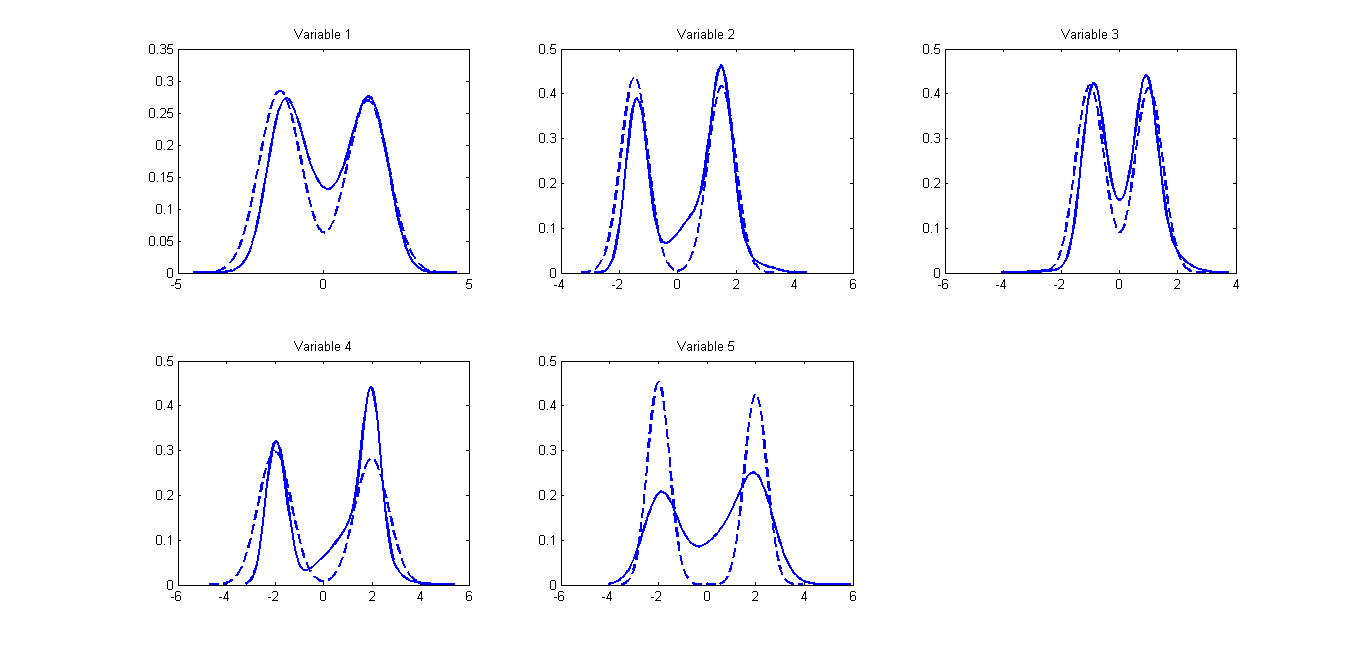}
\caption[]{Simulated example 2. Online VB approximate (solid lines) predictive densities of components of $\boldsymbol\beta_{n+1}$ and the associated true densities (dashed lines) for $K=261,144$ and $n=600$.}
\label{predicted600m5}
\end{figure}

We repeated the analysis with $n=600$ and $K=261,144$. The run time was 5.68 hours. The estimated predictive densities are shown in Figure~\ref{predicted600m5}. The estimates are similar to those given in Figure~\ref{predicted400}. 
    
\subsection{Simulated example 3: online VB and MCMC comparison}

This section compares online VB and MCMC estimates. In the first example (results not shown) we fitted the spatial hierarchical model of simulated example 2 with $n=400$ and $K=2,500$. The run time for the online VB algorithm was 5.33 minutes whereas one iteration of the MCMC algorithm took about 14 minutes. Because of the MCMC computational cost, in order to compare online VB and MCMC estimates we instead generated data from the standard hierarchical model 
\begin{equation}\textbf{Y}_{it}=\textbf{X}_{it}\boldsymbol\beta_{i}+\boldsymbol\epsilon_{it},\; i=1,...,n,\; t=1,...,T,\nonumber \end{equation}
with $K=2,500$, $n=400$ and $T=5$. $\boldsymbol\epsilon_{it}$, $\boldsymbol\beta_{i}$ and $\textbf{X}_{it}$ are defined as in simulated example 2. 

A simplified version of the MCMC algorithm described in supplemental appendix A and the online VB algorithm were applied to these data. The MCMC algorithm was ran with 5,000 iterations, with the first 1,000 iterations discarded as burn-in and every 5th of the remaining 4,000 iterations used for posterior summaries. The run time for the online VB and MCMC algorithms was 1.67 minutes and 19 hours, respectively. 

\begin{figure} 
\centering
\includegraphics[scale=0.52]{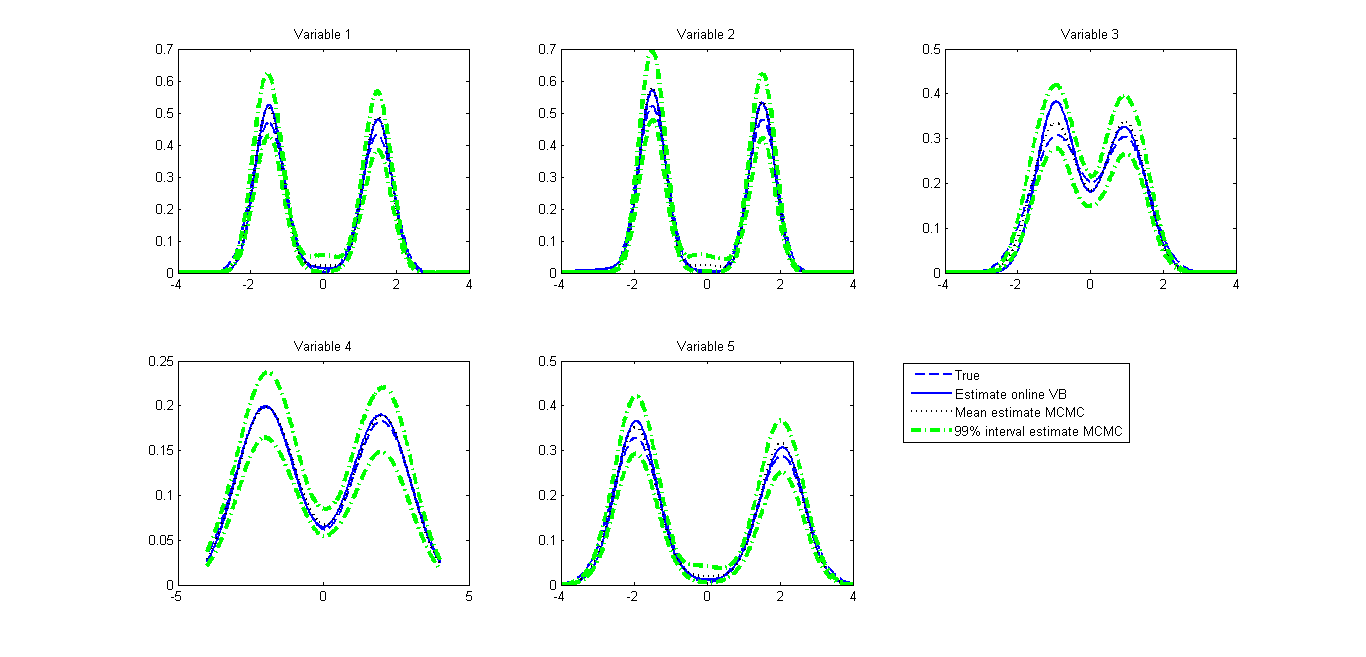}
\caption[]{Simulated example 3. True densities and predictive densities estimates from online VB and MCMC algorithm.}
\label{vbmcmc}
\end{figure}

Figure ~\ref{vbmcmc} shows the MCMC and online VB predictive densities for each of the regression coefficients. The two sets of estimates are indistinguishable.

\subsection{Effect of reordering subjects and discounting}

\begin{figure} 
\centering
\includegraphics[scale=0.5]{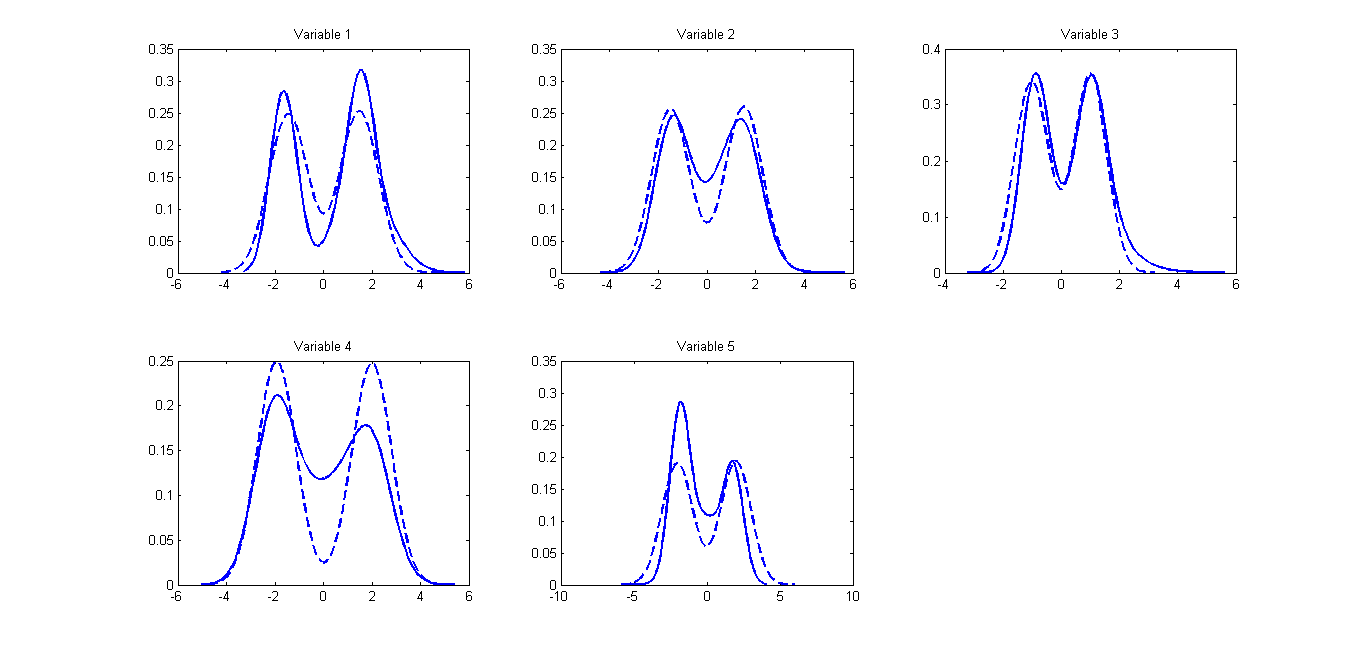}
\caption[]{Subjects are randomly ordered, same discounting as in no reordering case. Online VB approximate (solid lines) predictive densities of components of $\boldsymbol\beta_{n+1}$ and the associated true densities (dashed lines).}
\label{predicted400ordertest2}
\end{figure}

In this subsection we illustrate the importance of discounting in the online VB algorithm and the effect of the order in which individual data are read in. We focus on the setup of Simulation example 2. Figure ~\ref{predicted400ordertest2} plots the densities estimated with a random reordering of the subjects. Comparing with Figure ~\ref{predicted400}, reordering the subjects does not seem to affect the estimates of the shape and location of the regression parameter distributions. 

\begin{figure} 
\centering
\includegraphics[scale=0.5]{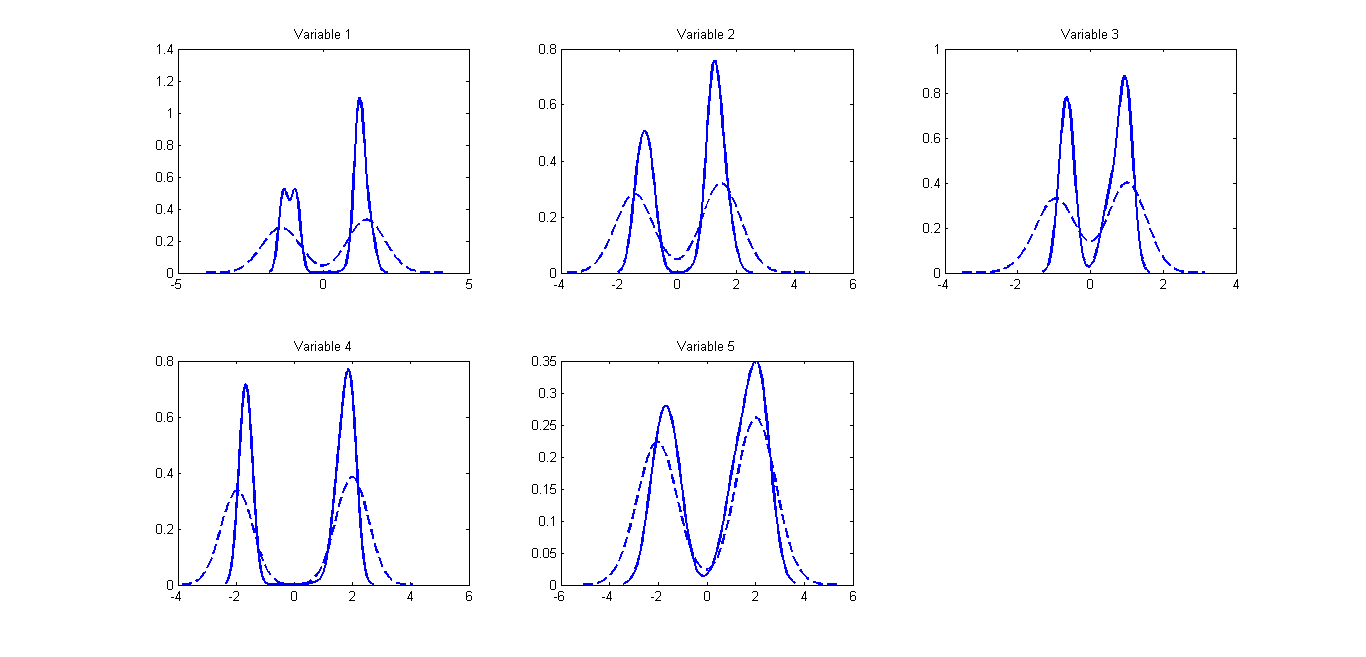}
\caption[]{Subjects data read in in the order 1 to n, no discounting. Online VB approximate (solid lines) predictive densities of components of $\boldsymbol\beta_{n+1}$ and the associated true densities (dashed lines).}
\label{predicted1nodiscounting}
\end{figure}

\begin{figure} 
\centering
\includegraphics[scale=0.5]{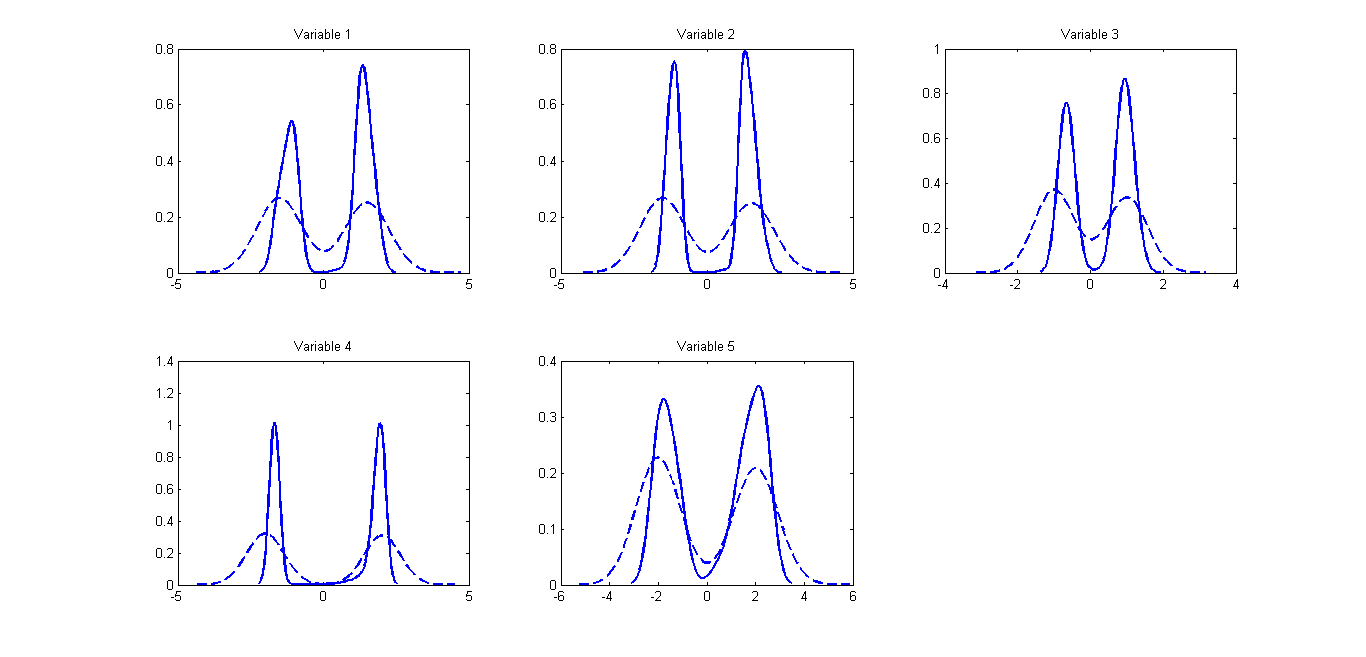}
\caption[]{Subjects are randomly ordered, no discounting. Online VB approximate (solid lines) predictive densities of components of $\boldsymbol\beta_{n+1}$ and the associated true densities (dashed lines).}
\label{predicted400ndordertest1}
\end{figure}

Figures ~\ref{predicted1nodiscounting} and ~\ref{predicted400ndordertest1} show the estimated distributions with and without reordering of the subjects' data respectively, with no discounting ($h(l)=0$). Ignoring discounting leads to poor estimates of the variance of the distributions. 

\section{APPLICATION TO QUANTITATIVE ANALYSIS OF MRI SIGNAL INTENSITY IN KNEE OSTEOARTHRITIS}

Osteoarthritis (OA) is the most common joint disorder and cause of disability in adults. This condition most often manifests itself in the form of pain and stiffness in the joints. Possible data for diagnosing OA and monitoring its progression over time include clinical indicators, x-rays, CT scans, and MRI scans. Unlike x-rays and CT scans, MRI images provide detailed three-dimensional views of soft tissue such as cartilage, muscle, ligaments and tendons, and are very useful for detecting early OA (\citealp{DCJ08}). 

The osteoarthritis initiative (OAI) (\href{http://oai.epi-ucsf.org}{oai.epi-ucsf.org}) conducted a study with 4796 men and women, aged 45-79 years, who either have or are at increased risk of developing knee OA. X-rays and MRI images of the left and right knee were taken at baseline visit, 12 month, 18 month, 24 month, or 36 month follow-ups and not all the 4796 subjects have data at all four time points.

There has been interest in comparing knee cartilage quantitative measures across knee compartments, time or participants (\citet{Carballidoetal10}, \citet{Balamoodyetal10}). \citet{Carballidoetal10} compared  compartment averaged mean $T_2$ laminar integrity over time for a handful of participants, using paired t-tests. We apply the model and estimation procedure of the previous Sections to analyze signal intensity instead of $T_2$ laminar integrity. Our main goal is to assess how signal intensity varies among subjects across knee compartments and over time. 

We selected a subset of 131 subjects from the OAI database, all of whom have images at each of the four visits: baseline, 12 month, 24 month, and 36 month. It is straightforward to allow different numbers of follow-up observations but we focus on subjects with complete data for simplicity. X-rays indicated signs of knee OA for 81 of the subjects at the baseline visit, 9 of whom had no pain at any of the three subsequent visits. 13 subjects neither had OA at the baseline visit nor pain at the three follow-up visits.    

Images consist of sagittal three-dimensional double echo in steady state (DESS), repetition time of 16.3 ms, echo time of 4.7 ms, bandwidth of 185 Hz/pixel, slice thickness of 0.7 mm, and in-plane spatial resolution of 0.365 mm $\times$ 0.365 mm. Data matrices are $384 \times 384 \times 160$. Each images was segmented and each pixel mapped to one of 3 structures: tibia, femur, and cartilage, using the seeded region growing segmentation tools in ImageJ (\href{http://rsb.info.nih.gov/ij/}{rsb.info.nih.gov/ij/}) on a slice-by-slice basis, for every fourth of the middle 50 slices. 

Before applying the methodology of the previous Sections we first used Wilcoxon Rank-Sum tests to compare the medians of the distributions of signal intensity for subjects with and without OA pain/signs across visits and knee compartments. The multimodality of the distributions of signal intensity among subjects at each time point and within each compartment, and the unequal sample size prevented the use of two-sample t-tests.  Results of the Wilcoxon Rank-Sum tests are shown in  Table~\ref{tab1}. There is a statistically significant difference between the distributions of signal intensity of subjects with OA pain/signs and the signal intensity of subjects with no OA pain/signs in the cartilage and tibia compartments at 24 and 36 month visits. In addition, the median signal intensity difference appears to be higher at 36-month follow-up visit compared to baseline visit across all three compartments, and within the cartilage compartment compared to femur and tibia compartments. 

\begin{table}[ht]
\caption{Comparison of median signal intensity for subjects with and without OA pain/signs across visits and knee compartments}\label{tab1}
\centering
\begin{tabular}{c c c c}
\hline
Variable & Median OA& Median no OA & One-sided p-value Wilcoxon rank sum test\\ 
\hline
Femur baseline  &	0.678 & 0.669&0.20\\
Femur 12 month       &	0.723 & 0.704&0.14\\
Femur 24 month        &  0.830 & 0.759&0.20\\
Femur 36 month       &  0.998 & 1.017&0.43\\
Tibia baseline  &	0.770 & 0.749&0.26\\
Tibia 12 month       &	0.767 & 0.741&0.14\\
Tibia 24 month        &  0.893 & 0.828&0.03\\
Tibia 36 month       &  1.092& 1.031&0.02\\
Cartilage baseline  &	2.310 & 2.232 &0.22\\
Cartilage 12 month       &	2.424 & 2.336 &0.30\\
Cartilage 24 month        &  2.676 & 2.591&0.08\\
Cartilage 36 month       &  3.276 & 3.058&0.01\\
\hline
\end{tabular}
\end{table}

\begin{figure} 
\centering
\includegraphics[scale=0.53]{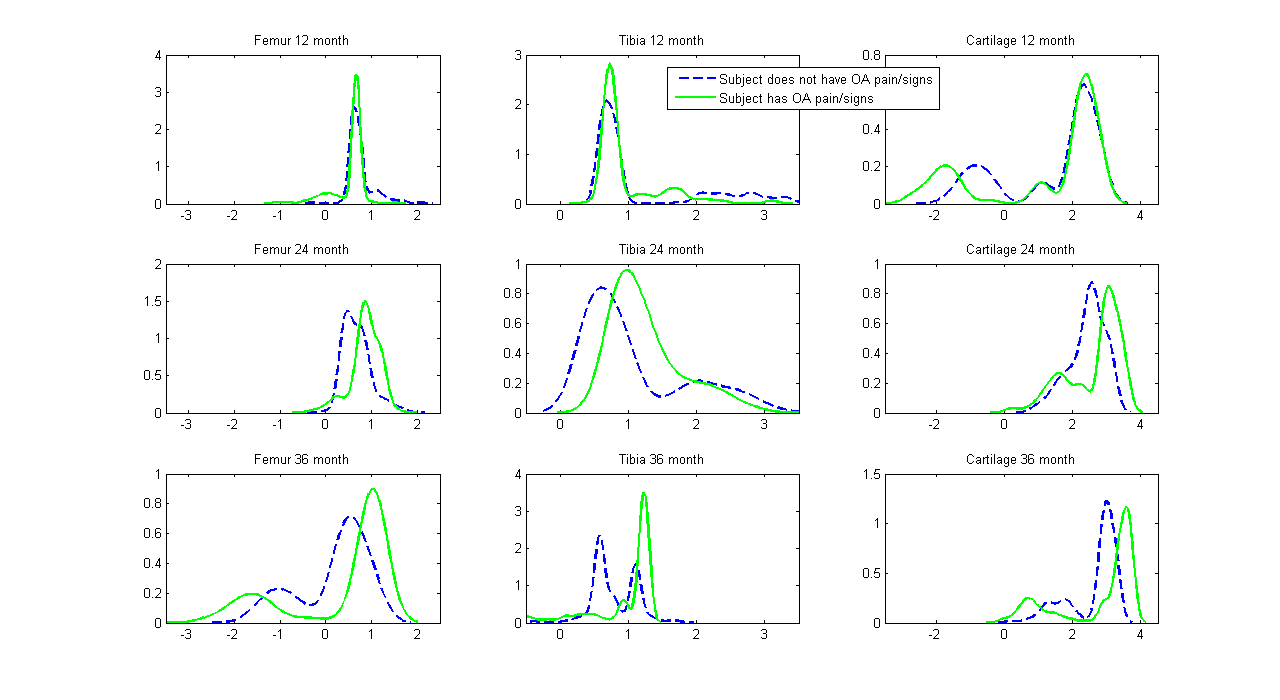}
\caption[]{Predictive densities of components of $\boldsymbol\beta_{n+1}$.}
\label{compartmentsm}
\end{figure}

The Wilcoxon rank sum test assumes that the distribution of signal intensity for subjects with OA pain/signs differs from that for the subjects with no OA pain/signs only with respect to the median (shapes and spreads of the distributions are identical). Also, the test is not aimed at characterizing the shape of the underlying distribution of the data. One may be interested in comparing the shapes of the distributions instead. To this end, we next fitted the hierarchical model given by (\ref{notemp}) to the data, with design matrix consisting of indicator variables for visits, knee compartments, and OA pain/signs. 
Figure~\ref{compartmentsm} shows the predictive distribution of  $\boldsymbol\beta_{n+1}$. Clearly, the two sets of distributions (subject has OA, subject does not have OA) have different shapes and spreads. The estimated distributions when the subject has OA pain/signs are bimodal or trimodal, with the location of the most prominent mode shifting to the right of that of the distribution when subject does not have OA pain/signs at 24 and 36 month visits, but not at 12 month visit. This indicates the subgroup of subjects with OA pain/signs that have higher signal intensity than those with not OA. 
Also, there is a subgroup of subjects with OA that have lower signal intensity than some subjects with no OA in the femur compartment at the 36 month visit, and in the cartilage compartment at 12 and 36 month visits.     
 
\section{DISCUSSION}

In this paper we have proposed an online variational Bayes algorithm for estimation and inference in flexible hierarchical regression models for correlated high-dimensional data. The methodology was illustrated first via simulated examples and then using knee MRI data from the Osteoarthritis Initiative, and was shown to produce good results in both types of data. The online variational Bayes algorithm is developped for hierarchical regression models but it can be adapted to various classes of models for high-dimensional data. 

\section*{ACKNOWLEDGMENTS}

The OAI is a public-private partnership comprised of five contracts (N01-AR-2-2258; N01-AR-2-2259; N01-AR-2-2260; N01-AR-2-2261; N01-AR-2-2262) funded by the National Institutes of Health (NIH) and conducted by the OAI Study Investigators. Private funding partners include Merck Research Laboratories; Novartis Pharmaceuticals Corporation, GlaxoSmithKline; and Pfizer, Inc. Private sector funding for the OAI is managed by the Foundation for the NIH. This manuscript was prepared using an OAI public use data set and does not necessarily reflect the opinions or views of the OAI investigators, the NIH, or the private funding partners.

This research was partially supported by grant R01 ES017240-01 from the National Institute of Environmental Health Sciences (NIEHS) of the NIH and the Statistics and Applied Mathematical Sciences Institute (SAMSI) program on the Analysis of Object Data. Morris' effort was supported by the National Cancer Institute (CA-107304).   

\section*{APPENDIX} 

\appendix

\section{MCMC CONDITIONAL POSTERIOR DISTRIBUTIONS}\label{MCMC}
\begin{itemize}
\item[] \textit{Step 1.} Sample the indicator variables $z_i$ from
\begin{equation}\Pr(z_i=r)\propto exp\left\{-0.5\sigma^2\sum_t(\mathbf{Y}_{it}-\boldsymbol\eta_{i}\boldsymbol\mu_{it}-\mathbf{X}_{it}\boldsymbol\beta_r^{*})'(\mathbf{Y}_{it}-\boldsymbol\eta_{i}\boldsymbol\mu_{it}-\mathbf{X}_{it}\boldsymbol\beta_r^{*})\right\}\nonumber \end{equation}
\item[] \textit{Step 2.} Sample the component parameters $\theta_r^{*}$ and $\boldsymbol\beta_r^{*}$ from 
\begin{equation}\theta_r^{*}\sim Ga\left(a_{\theta}+0.5T\sum_{i:z_i=r}z_i, b_{\theta}+0.5\sum_{t,i: z_i=r} (\boldsymbol\mu_{it}-\boldsymbol\mu_{i,t-1})'(\boldsymbol\mu_{it}-\boldsymbol\mu_{i,t-1})\right)\nonumber \end{equation}
\begin{equation}\boldsymbol\beta_r^{*}\sim N\left(\hat{\boldsymbol\Sigma}_r\left(\boldsymbol\beta_{0}'\Sigma_0^{-1}+\sigma^{-2}\sum_{t,i:z_i=r}(\mathbf{Y}_{it}-\boldsymbol\eta_{i}\boldsymbol\mu_{it})'\mathbf{X}_{it}\right),\hat{\boldsymbol\Sigma}_r =\left(\Sigma_0^{-1}+\sigma^{-2}\sum_{t,i:z_i=r}\mathbf{X}_{it}'\mathbf{X}_{it}\right)^{-1}\right)\nonumber \end{equation}
\item[] \textit{Step 3.} Sample the weight parameters $v_r$ from 
\begin{equation}v_r\sim Be\left(1+\sum_{i: z_i=r} z_i,\alpha+\sum_{i:z_i=r} \sum_{l=r+1}^R z_i\right)\nonumber \end{equation}
\item[] \textit{Step 4.} Sample the precision parameter $\alpha$ from 
\begin{equation}\alpha \sim Ga\left(a_{\alpha}+\sum_{i: z_i=r}z_i,b_{\alpha}+\sum_{i: z_i>r}z_i\right)\nonumber \end{equation}
\item[] \textit{Step 5.} Sample the common factors $\boldsymbol\mu_{it}$ from 
\begin{equation}\boldsymbol\mu_{i0}\sim N\left(\left(\vartheta^{-1}+\theta_i\right)^{-1}\left(\boldsymbol\mu_{i1}\theta_i+\boldsymbol\mu_{0}\vartheta^{-1} \right),\left(\vartheta^{-1}+\theta_i\right)^{-1}\right)\nonumber \end{equation} 
\begin{equation}\boldsymbol\mu_{it}\sim N\left(\left(\sigma^2 \boldsymbol\eta_i'\boldsymbol\eta_i+ \theta_i\right)^{-1}\left(\sigma^2\boldsymbol\eta_i(\mathbf{Y}_{it}-\mathbf{X}_{it}\boldsymbol\beta_i)+\boldsymbol\mu_{i,t-1} \theta_i\right),\left(\sigma^2 \boldsymbol\eta_i'\boldsymbol\eta_i+ \theta_i\right)^{-1}\right)\nonumber \end{equation} 
\item[] \textit{Step 6.} Sample the loadings $\boldsymbol\eta_{ij}$ from 
\begin{equation}\boldsymbol\eta_{ij}\sim N\left(\boldsymbol\Psi_{ij}\left(\sigma^2\sum_t\mu_{it,j}(\mathbf{Y}_{it}-\mathbf{X}_{it}\boldsymbol\beta_i)\right),\boldsymbol\Psi_{i}=\left(K\sigma^2\sum_t\mu_{it,j}^2\mathbf{I}+ \tau^{-1}(I-\rho C)^{-1}\Omega\right)^{-1}\right)\nonumber \end{equation}
\item[] \textit{Step 7.} Sample the parameter $\tau$ from
\begin{equation}\tau\sim Ga\left(a_{\tau}+0.5mnK,b_{\tau}+0.5\sum_{i,j}(\boldsymbol\eta_{ij}'\Omega^{-1}(I-\rho C)\boldsymbol\eta_{ij} \right)\nonumber \end{equation}
\item[] \textit{Step 8.} Sample the parameter $\rho$ from
\begin{equation}\Pr(\rho=\rho_l) \propto exp\left\{-0.5\sum_{i,j}\boldsymbol\eta_{ij}'\Omega^{-1}(I-\rho_l C)\boldsymbol\eta_{ij}\right\}\nonumber \end{equation}
\item[] \textit{Step 9.} Sample the inverse variance $\sigma^2$ from 
\begin{equation}\sigma^{2} \sim Ga\left(a_{\sigma}+nKT/2,b_{\sigma}+\frac{1}{2}\sum_{t,i: z_i=r}\left\{(\mathbf{Y}_{it}-\boldsymbol\eta_{i}\boldsymbol\mu_{it}-\mathbf{X}_{it}\tilde{\boldsymbol\beta}_{i})'(\mathbf{Y}_{it}-\boldsymbol\eta_{i}\boldsymbol\mu_{it}-\mathbf{X}_{it}\tilde{\boldsymbol\beta}_{i})\right\}\right)\nonumber \end{equation}
\end{itemize}

\section{VARIATIONAL OBJECTIVE FUNCTION}\label{objectiveVB}

Let $\mathbf{W}=(V,\boldsymbol\Theta^{*},Z,\boldsymbol\eta,\boldsymbol\mu,\rho,\tau,\sigma^2,\alpha)$ denote the vector of unknown parameters and $A=(a_{\alpha},b_{\alpha},a_{\theta},b_{\theta},a_{\tau},b_{\tau},\boldsymbol\mu_0,\vartheta)$ the known hyper-parameters.

The posterior distribution $p(\mathbf{W}|\mathbf{Y},\mathbf{X},A)$ is approximated with a more tractable distribution $q(\mathbf{W})$ which maximizes the lower bound on the log marginal likelihood given by:
\begin{eqnarray}
\ell(\mathbf{Y}|\mathbf{X},A) &=& \int q(\mathbf{W})\;log\frac{p(\mathbf{Y}|\mathbf{W},\mathbf{X},A)p(\mathbf{W}|A)}{q(\mathbf{W})}d\mathbf{W} \nonumber \\
&=& E_q\left[log\;p(\sigma^2)-log\;q(\sigma^2)\right]+E_q\left[log\;p(\alpha)-log\;q(\alpha)\right]\nonumber \\
&+& E_q\left[log\;p(\rho)-log\;q(\rho)\right]+E_q\left[log\;p(\tau)-log\;q(\tau)\right]\nonumber \\
&+& E_q\left[log\;p(V|\alpha)-log\;q(V)\right] +E_q\left[log\;p(\boldsymbol\theta^{*})-log\;q(\boldsymbol\theta^{*})\right] +E_q\left[log\;p(\boldsymbol\beta^{*})-log\;q(\boldsymbol\beta^{*})\right]\nonumber \\
&+& E_q\left[log\;p(Z|V)-log\;q(Z)\right]+E_q\left[log\;p(\boldsymbol\mu|Z)-log\;q(\boldsymbol\mu)\right]+E_q\left[log\;p(\boldsymbol\eta)-log\;q(\boldsymbol\eta)\right]\nonumber \\
&+& E_q\left[log\;p(\mathbf{Y}|\mathbf{W},\mathbf{X},A)\right]. \label{llb}
\end{eqnarray}

The expectations in the objective function are evaluated as
\begin{equation}E_q\left[log\;p(\sigma^2)-log\;q(\sigma^2)\right] =(\psi(\tilde{a}_{\sigma})-log(\tilde{b}_{\sigma}))(a_{\sigma}-\tilde{a}_{\sigma})- \frac{\tilde{a}_{\sigma}}{\tilde{b}_{\sigma}}(b_{\sigma}-\tilde{b}_{\sigma})+log\left(\frac{b_{\sigma}^{a_{\sigma}} \Gamma(\tilde{a}_{\sigma})}{\tilde{b}_{\sigma}^{\tilde{a}_{\sigma}} \Gamma(a_{\sigma})}\right).\nonumber \end{equation}
\begin{equation}E_q\left[log\;p(\alpha)-log\;q(\alpha)\right] =(\psi(\tilde{a}_{\alpha})-log(\tilde{b}_{\alpha}))(a_{\alpha}-\tilde{a}_{\alpha})- \frac{\tilde{a}_{\alpha}}{\tilde{b}_{\alpha}}(b_{\alpha}-\tilde{b}_{\alpha})+log\left(\frac{b_{\alpha}^{a_{\alpha}} \Gamma(\tilde{a}_{\alpha})}{\tilde{b}_{\alpha}^{\tilde{a}_{\alpha}} \Gamma(a_{\alpha})}\right).\nonumber \end{equation}
\begin{equation}E_q\left[log\;p(\tau)-log\;q(\tau)\right] =(\psi(\tilde{a}_{\tau})-log(\tilde{b}_{\tau}))(a_{\tau}-\tilde{a}_{\tau})- \frac{\tilde{a}_{\tau}}{\tilde{b}_{\tau}}(b_{\tau}-\tilde{b}_{\tau})+log\left(\frac{b_{\tau}^{a_{\tau}} \Gamma(\tilde{a}_{\tau})}{\tilde{b}_{\tau}^{\tilde{a}_{\tau}} \Gamma(a_{\tau})}\right).\nonumber \end{equation}
\begin{eqnarray*}
\lefteqn{E_q\left[log\;p(V|\alpha)-log\;q(V)\right] =}\\
 & & \left(\frac{\tilde{a}_{\alpha}}{\tilde{b}_{\alpha}}-1\right)\left[\psi(\gamma_{r2})-\psi(\gamma_{r2}+\gamma_{r2})\right]+\psi(\tilde{a}_{\alpha})-log(\tilde{b}_{\alpha})-\left[\psi(\gamma_{r2})-\psi(\gamma_{r1}+\gamma_{r2})\right].
\end{eqnarray*}
\begin{equation}
E_q\left[log\;p(\theta^{*})-log\;q(\theta^{*})\right] = \sum_{r=1}^R \left\{(\psi(\tau_{r1})-log(\tau_{r2}))(a_{\tau}-\tau_{r1}) -\frac{\tau_{r1}}{\tau_{r2}}(b_{\tau}-\tau_{r2})+log\left(\frac{b_{\tau}^{a_{\tau}} \Gamma(\tau_{r1})}{\tau_{r2}^{\tau_{r1}} \Gamma(a_{\tau})}\right)\right\}. \nonumber \end{equation}
\begin{equation}
E_q\left[log\;p(Z|V)-log\;q(Z)\right] = \sum_{i,t,r}\kappa_{ir}\left\{\psi(\gamma_{r1})-\psi(\gamma_{r1}+\gamma_{r2})+\sum_{l=1}^{r-1}\left[\psi(\gamma_{l2})-\psi(\gamma_{l1}+\gamma_{l2})\right] -log(\kappa_{ir})\right\}.\nonumber\end{equation}
\begin{dmath*}
E_q\left[log\;p(\boldsymbol\mu|Z)-log\;q(\boldsymbol\mu)\right] = 
 \frac{1}{2}\sum_{i=1}^n \sum_{t=1}^T \sum_{r=1}^R \kappa_{ir}\left\{ \left[\psi(\tau_{r1})-log(\tau_{r2})-log(2\pi)\right]- \left[(\boldsymbol\lambda_{it,1}-\boldsymbol\lambda_{i,t-1,1})'(\boldsymbol\lambda_{it,1}-\boldsymbol\lambda_{i,t-1,1})+tr(\boldsymbol\lambda_{it,2}+\boldsymbol\lambda_{i,t-1,2})\right]\right\}-
 \frac{nT}{2}\left(log(\left|\boldsymbol\lambda_{it,2}\right|)+log(2\pi)+1\right).
\end{dmath*}
\begin{dmath*}
E_q\left[log\;p(\boldsymbol\eta_{i})-log\;q(\boldsymbol\eta_{i})\right] = -0.5m\:log(\left|\boldsymbol\Psi_{i}\right|)
+ 0.5m\left(K(\psi(\tilde{a}_{\tau})-log(\tilde{b}_{\tau}))-\sum_{l=1}^{M+1}\pi_{l}log\left(\left|(I-\rho_l C)^{-1}\Omega\right|\right)\right)  
- \frac{\tilde{a}_{\tau}}{2\tilde{b}_{\tau}}\sum_{l=1}^{M+1}\left\{m\;tr\left(\Omega^{-1}(I-\rho_l C)\boldsymbol\Psi_{i}\right)+\sum_{k=1}^m\boldsymbol\xi_{ik}'\Omega^{-1}(I-\rho_l C)\boldsymbol\xi_{ik}\right\}.
\end{dmath*}
\begin{dmath*}
E_q\left[log\;p(\mathbf{Y}|\mathbf{W},\mathbf{X},A)\right] = 
 \frac{1}{2}\frac{\tilde{a}_{\sigma}}{\tilde{b}_{\sigma}}\sum_{i,t,r}\kappa_{ir}\left( (\mathbf{Y}_{it}-\boldsymbol\xi_{i}\boldsymbol\lambda_{it,1}-\mathbf{X}_{it}\tilde{\boldsymbol\beta}_{0r})'(\mathbf{Y}_{it}-\boldsymbol\xi_{i}\boldsymbol\lambda_{it,1}-\mathbf{X}_{it}\tilde{\boldsymbol\beta}_{0r})
+(tr(\boldsymbol\Psi_{i})+\boldsymbol\xi_{i}'\boldsymbol\xi_{i})tr(\boldsymbol\lambda_{it,2})+\lambda_{it,1}'tr(\boldsymbol\Psi_{i})\boldsymbol\lambda_{it,1}+ tr\left(\mathbf{X}_{it}'\mathbf{X}_{it}\tilde{\Sigma}_{0r}\right)\right)+
 \frac{nT}{2}\left(\psi(\tilde{a}_{\sigma})-log(\tilde{b}_{\sigma})-log(2\pi)\right). 
\end{dmath*}

\section{VARIATIONAL BAYES UPDATE EQUATIONS}\label{updateequationVB}

The variational Bayes update equations are derived as: 
\begin{itemize}
\item $\tilde{a}_{\sigma}=a_{\sigma}+nKT/2$\\
      $\tilde{b}_{\sigma}= b_{\sigma}+\frac{1}{2}\sum_{i,t,r}\kappa_{ir}\left\{(\mathbf{Y}_{it}-\boldsymbol\xi_{i}\boldsymbol\lambda_{it,1}-\mathbf{X}_{it}\tilde{\boldsymbol\beta}_{0r})'(\mathbf{Y}_{it}-\boldsymbol\xi_{i}\boldsymbol\lambda_{it,1}-\mathbf{X}_{it}\tilde{\boldsymbol\beta}_{0r})\right\}$\\
$+\frac{1}{2}\sum_{i,t,r}\kappa_{ir}\left\{(tr(\boldsymbol\Psi_{i})+\boldsymbol\xi_{i}'\boldsymbol\xi_{i})tr(\boldsymbol\lambda_{it,2})+\lambda_{it,1}'tr(\boldsymbol\Psi_{i})\boldsymbol\lambda_{it,1}+ tr\left(\mathbf{X}_{it}'\mathbf{X}_{it}\tilde{\Sigma}_{0r}\right)\right\}$.
\item $\tilde{a}_{\alpha}=a_{\alpha}+R-1, \; \tilde{b}_{\alpha}=b_{\alpha}-\sum_{r=1}^{R-1} (\psi(\gamma_{r2})-\psi(\gamma_{r1}+\gamma_{r2})).$
\item $\tilde{a}_{\theta_r}=a_{\theta}+0.5T\sum_{i}\kappa_{ir}$,\\
      $\tilde{b}_{\theta_r}=b_{\theta}+0.5\sum_{i,t} \kappa_{ir}\left\{(\boldsymbol\lambda_{it,1}-\boldsymbol\lambda_{i,t-1,1})'(\boldsymbol\lambda_{it,1}-\boldsymbol\lambda_{i,t-1,1})+tr(\boldsymbol\lambda_{it,2}+\boldsymbol\lambda_{i,t-1,2})\right\}$.
\item $\tilde{\Sigma}_{0r} =\left(\Sigma_0^{-1}+\frac{\tilde{a}_{\sigma}}{\tilde{b}_{\sigma}}\sum_i\sum_t\kappa_{ir}\mathbf{X}_{it}'\mathbf{X}_{it}\right)^{-1}$,\\
		$\tilde{\boldsymbol\beta}_{0r}=\tilde{\Sigma}_{0r} \left(\boldsymbol\beta_{0}'\Sigma_0^{-1}+\frac{\tilde{a}_{\sigma}}{\tilde{b}_{\sigma}}\sum_i\sum_t\kappa_{ir}(\mathbf{Y}_{it}-\boldsymbol\xi_{i}\boldsymbol\lambda_{it,1})'\mathbf{X}_{it}\right).$   
\item $\boldsymbol\lambda_{i0,2}=\left(\vartheta^{-1}+\sum_{r} \kappa_{ir} \frac{\tilde{a}_{\theta_r}}{\tilde{b}_{\theta_r}}\right)^{-1},$ 
$\boldsymbol\lambda_{i0,1}=\boldsymbol\lambda_{i0,2}\left(\boldsymbol\lambda_{i1,1} \sum_{r} \kappa_{ir} \frac{\tilde{a}_{\theta_r}}{\tilde{b}_{\theta_r}}+\boldsymbol\mu_0\vartheta^{-1}\right),$\\
$\boldsymbol\lambda_{it,2}=\left((tr(\boldsymbol\Psi_{i})+\boldsymbol\xi_{i}'\boldsymbol\xi_{i})\frac{\tilde{a}_{\sigma}}{\tilde{b}_{\sigma}}+\sum_{r} \kappa_{ir} \frac{\tilde{a}_{\theta_r}}{\tilde{b}_{\theta_r}}\right)^{-1},$   \\
$\boldsymbol\lambda_{it,1}=\boldsymbol\lambda_{it,2}\sum_{r}\ \kappa_{ir}\left(\boldsymbol\xi_{i}'(\mathbf{Y}_{it}-\mathbf{X}_{it}\boldsymbol\beta_{0r})\frac{\tilde{a}_{\sigma}}{\tilde{b}_{\sigma}}+\boldsymbol\lambda_{i,t-1,1}\frac{\tilde{a}_{\theta_r}}{\tilde{b}_{\theta_r}}\right).$ 
\item $\gamma_{r1} =1+\sum_{i=1}^n \kappa_{ir}, \; \gamma_{r2} =\frac{\tilde{a}_{\alpha}}{\tilde{b}_{\alpha}}+\sum_{i=1}^n \sum_{l=r+1}^R \kappa_{il}.$
\item $\psi_{ik}=\left(T\frac{\tilde{a}_{\sigma}}{\tilde{b}_{\sigma}}+\frac{\tilde{a}_{\tau}}{\tilde{b}_{\tau}}\Omega_{kk}^{-1}\right)^{-1}$, $k=1,...,K$,\\
      $\xi_{ijk}=\psi_{ik}\frac{\tilde{a}_{\sigma}}{\tilde{b}_{\sigma}}\sum_r\kappa_{ir}\sum_t \lambda_{ikt,1}(\mathbf{Y}_{ikt}-\mathbf{X}_{ikt}\tilde{\boldsymbol\beta}_{0r})$, $j=1,...,m$, $k=1,...,K$
\item $\tilde{a}_{\tau}=a_{\tau}+0.5mnK$, \\
      $\tilde{b}_{\tau}=b_{\tau}+0.5\sum_i\sum_l \pi_{l}\left\{m\:tr(\Omega^{-1}\boldsymbol\Psi_{i})+\sum_{l=1}^m\boldsymbol\xi_{il}'\Omega^{-1}(I-\rho_l C)\boldsymbol\xi_{il}\right\}.$
\item $\kappa_{ir}\propto exp(w_{ir})$,\\ $w_{ir}=-\frac{\tilde{a}_{\sigma}}{2\tilde{b}_{\sigma}}\sum_t\left\{(\mathbf{Y}_{it}-\boldsymbol\xi_{i}\boldsymbol\lambda_{it,1}-\mathbf{X}_{it}\tilde{\boldsymbol\beta}_{0r})'(\mathbf{Y}_{it}-\boldsymbol\xi_{i}\boldsymbol\lambda_{it,1}-\mathbf{X}_{it}\tilde{\boldsymbol\beta}_{0r})\right\}$ \\$-$ $\frac{\tilde{a}_{\sigma}}{2\tilde{b}_{\sigma}}\sum_t\left\{(tr(\boldsymbol\Psi_{i})+\boldsymbol\xi_{i}'\boldsymbol\xi_{i})tr(\boldsymbol\lambda_{it,2})+\lambda_{it,1}'tr(\boldsymbol\Psi_{i})\boldsymbol\lambda_{it,1}+ tr\left(\mathbf{X}_{it}'\mathbf{X}_{it}\tilde{\Sigma}_{0r}\right)\right\}$ \\$-$ $0.5\frac{\tilde{a}_{\theta_r}}{\tilde{b}_{\theta_r}}\sum_t\left\{(\boldsymbol\lambda_{it,1}-\boldsymbol\lambda_{i,t-1,1})'(\boldsymbol\lambda_{it,1}-\boldsymbol\lambda_{i,t-1,1})+tr(\boldsymbol\lambda_{it,2}+\boldsymbol\lambda_{i,t-1,2})\right\}$ \\$+$ $\psi(\gamma_{r1})-\psi(\gamma_{r1}+\gamma_{r2})+\sum_{l=1}^{r-1} \left\{\psi(\gamma_{l2})-\psi(\gamma_{l1}+\gamma_{l2})\right\}$ $+$ $\frac{KT}{2}(\psi(\tilde{a}_{\sigma})-log(\tilde{b}_{\sigma})),$ \\where $\psi(.)$ is the digamma function.
\item $\tilde{\pi}_{l} \propto exp(\varpi_{l})$,
      $\varpi_{l}=0.5m\:n\:K\:(\psi(\tilde{a}_{\tau})-log(\tilde{b}_{\tau})) -0.5m\:n\: (log(\left|\Omega\right|)- log\left(\left|(I-\rho_{l} C) \right|\right))$\\  
      $-\frac{a_{\tau}}{2b_{\tau}}\sum_i \left\{m\:tr(\Omega^{-1}\boldsymbol\Psi_{i})+\sum_{k=1}^m\boldsymbol\xi_{ik}'\Omega^{-1}(I-\rho_l C)\boldsymbol\xi_{ik}\right\}.$\\
\end{itemize}


\begin{thebibliography}{}

\bibitem[Ansari, Jedidi and Dube(2002)]{AJD02}
Ansari, A., Jedidi, K., and Dube, L. (2002), ``Heterogeneous factor analysis models: a Bayesian approach," \textit{Psychometrika}, 67(1), 49--78.

\bibitem[Balamoody et al.(2010)]{Balamoodyetal10}
Balamoody, S., Williams, T. G., Waterton, J. C., Bowes, M., Hodgson, R., Taylor, C. J., and Hutchinson, C. E.  (2010), ``Comparison of 3T MR scanners in regional cartilage-thickness analysis in osteoarthritis: a cross-sectional multicenter, multivendor study," \textit{Arthritis Research \& Therapy}, 12:R202.
 
\bibitem[Banerjee et al.(2008)]{Banerjeeetal08}
Banerjee, S., Gelfand, A. E., Finley, A. O., and Sang, H. (2008), ``Gaussian predictive process models for
large spatial data sets," \textit{Journal of the Royal Statistical Society Serie B}, 70, 825–-848.

\bibitem[Banerjee, Dunson and Tokdar(2011)]{BDT11}
Banerjee, A., Dunson, D. B., and Tokdar, S. (2011), ``Efficient Gaussian Process Regression for Large Data Sets, Technical report, Duke University Department of Statistical Science. 

\bibitem[Barry and Pace(1999)]{BPa99}
Barry, R. , and Pace, R. K. (1999), ``A Monte Carlo Estimator of the Log Determinant of Large Sparse Matrices," \textit{Linear Algebra and its Applications}, 289, 41–-54.

\bibitem[Blei and Jordan(2006)]{BJ06}
Blei, D. M., and Jordan, M. I. (2006), ``Variational Inference for Dirichlet Process Mixtures,"  \textit{Bayesian Analysis}, 1(1), 121--144.

\bibitem[Bowman et al.(2008)]{Bowmanetal08}
Bowman, F. D., Caffo, B., Bassett, S. S., and Kilts, C. (2008), ``A Bayesian Hierarchical Framework for Spatial Modeling of fMRI Data," \textit{Neuroimage}, 39, 146--156.

\bibitem[Carballido-Gamio et al.(2010)]{Carballidoetal10}
Carballido-Gamio, J., Blumenkrantz, G., Lynch, J. A., Link, T. M., and Majumdar, S. (2010), ``Longitudinal analysis of MRI T2 knee cartilage laminar organization in a subset of patients from the osteoarthritis initiative," \textit{Magnetic Resonance in Medicine}, 63, 465--472.

\bibitem[Carvalho et al.(2010)]{Carvalhoetal10}
Carvalho, C. M., Lopes, H. F., Polson, N. G., and Taddy, M. A. (2010), ``Particle Learning for General Mixtures," \textit{Bayesian Analysis}, 5(4), 709--740.

\bibitem[Cheng et al.(2005)]{Chengetal05}
Cheng, L., Jiao, F., Schuurmans, D., and Wang, S. (2005), ``Variational Bayesian image modelling," In \textit{International Conference on Principles of Knowledge Representation and Reasoning}. 

\bibitem[Chopin et al.(2010)]{Chopinetal10}
Chopin, N., Iacobucci, A., Marin, J., Mengersen, K., Robert, C. P., Ryder, R. and Schafer, C. (2010), ``On particle learning,"  ArXiv e-prints URL: http://arxiv.org/abs/1006.0554.

\bibitem[Cressie and  Huang(1999)]{CH99}
Cressie, N., and Huang, H. (1999), ``Classes of Nonseparable, Spatio-Temporal Stationary Covariance Functions," J\textit{ournal of the American Statistical Association}, 94(448), 1330--1340.

%\bibitem[Cressie and  Johannesson(2008)]{CresJ08}
%Cressie, N. and Johannesson, G. (2008). ``Fixed rank Kriging for very large data sets," \textit{Journal of the Royal Statistical Society Serie B,} 70, 209–226.

\bibitem[Derado, Bowman and Kilts(2010)]{DBK10}
Derado, G., Bowman, F. B., and Kilts, C. D. (2010), ``Modeling the Spatial and Temporal Dependence in fMRI Data," \textit{Biometrics}, 66, 949–-957.

\bibitem[Ding, Cicuttini and Jones(2008)]{DCJ08}
Ding, C., Cicuttini, F., and Jones, G. (2010), ``How important is MRI for detecting early osteoarthritis?," \textit{Nature Clinical Practice Rheumatology  }, 4, 4--5.

\bibitem[Fearnhead(2004)]{Fear04}
Fearnhead, P. (2004), ``Particle filters for mixture models with an unknown number of components,"  \textit{Journal of Statistics and Computing}, 14, 11--21.

\bibitem[Gelfand and Vounatsou(2003)]{GV03}
Gelfand, A. E., and Vounatsou, P. (2003), ``Proper Multivariate Conditional Autoregressive Models for Spatial Data Analysis,"  \textit{Biostatistics}, 4(1), 11--25.

\bibitem[Gomes, Welling and Perona(2008)]{GWP08}
Gomes, R.,Welling, M., and Perona, P. (2008), ``Incremental learning of nonparametric Bayesian mixture models," In \textit{IEEE Conference on Computer
Vision and Pattern Recognition}.

\bibitem[Harrison and Green(2010)]{HG10}
Harrison, L. M., and Green, G.G.R. (2010), ``A Bayesian spatiotemporal model for very large data sets," \textit{NeuroImage}, 50, 1126–-1141.

\bibitem[Hoffman, Blei and Bach(2010)]{HBB10}
Hoffman, M. D., Blei, D. M., Bach, F. (2010), ``Online Learning for Latent Dirichlet Allocation," In \textit{Neural Information Processing Systems}.

\bibitem[Honkela and Valpola(2003)]{HV03}
Honkela, A. and Valpola, H. (2003), ``On-line variational Bayesian learning," In \textit{Proceedings of the 4th International Symposium on Independent Component Analysis and Blind Signal Separation (ICA '03)}, 803--808.

\bibitem[Hrafnkelsson and Cressie(2003)]{HC03}
Hrafnkelsson, B., and Cressie, N. (2003), ``Hierarchical Modeling of Count Data with Application to nuclear fall-out,"  \textit{Environmental and ecological Statistics}, 10, 179--200.

\bibitem[Kottas, Duan and Gelfand(2008)]{KDG08}
Kottas A., Duan J. A., and Gelfand, A. E. (2008), ``Modeling disease incidence data with spatial and spatio temporal Dirichlet process mixtures," \textit{Biometrical Journal}, 50(1), 29--42.

\bibitem[Jordan et al.(1999)]{Jordanetal99}
Jordan, M.,  Ghahramani, Z., Jaakkola, T., and Saul, L. (1999), ``An introduction to variational methods for graphical models," \textit{Machine Learning}, 37, 183–-233.

\bibitem[Lopes, Salazar and Gamerman(2008)]{LSG08}
Lopes, H. F., Salazar, E., and Gamerman, D. (2008), ``Spatial Dynamic Factor Analysis," \textit{Bayesian Analysis}, 3(4), 759--792.

\bibitem[Lopes et al.(2010)]{Lopesetal10}
Lopes, H., Carvalho, C. M.,  Johannes, M., and Polson, N. (2010), ``Particle Learning for Sequential Bayesian Computation (with discussion)," In J. Bernardo, M. J. Bayarri,J. Berger, A. Dawid, D. Heckerman, A. F. M. Smith and M. West (Eds.), \textit{Bayesian Statistics}, Volume 9. Oxford. In Press.

\bibitem[Morris and Carroll(2006)]{MC06}
Morris, J. S. and Carroll, R. J. (2006), ``Wavelet-based functional mixed models," \textit{Journal of the Royal Statistical Society, Serie B}, 68, 179--199.

\bibitem[Morris et al.(2011)]{Morrisetal11}
Morris, J. S., Baladandayuthapani, V., Herrick, R. C., Sanna, P., and Gutstein, H. G.  (2011),  Automated analysis of quantitative image data using isomorphic functional mixed models, with application to proteomics data.  \textit{The Annals of Applied Statistics}, 5(2A), 894--923.

\bibitem[Oikonomou, Tripoliti and Fotiadis(2010)]{OTF10}
Oikonomou, V.P.,   Tripoliti, E.E., and   Fotiadis, D.I. (2010), ``Bayesian Methods for fMRI Time-Series Analysis Using a Nonstationary Model for the Noise,"  \textit{IEEE Transactions, Information Technology in Biomedicine},  14(3), 664--674.  

\bibitem[Park et al.(2009)]{Parketal09}
Park, B. U., Mammen, E., Hardle, W., and Borak, S. (2009), ``Time Series Modelling With Semiparametric Factor Dynamics," \textit{Journal of the American Statistical Association}, 104, 284--298. 

\bibitem[Penny, Kiebel and Friston(2003)]{PKF03}
Penny, W., Kiebel, S., and Friston, K. (2003), ``Variational Bayesian inference for fMRI time series," \textit{NeuroImage}, 19, 727–-741.

\bibitem[Qi et al.(2008)]{Qietal08}
Qi, Y., Liu, D., Carin, L., and Dunson, D. (2008), ``Multi-task compressive sensing with Dirichlet process priors," \textit{International Conference on Machine Learning}.

\bibitem[Rue, Martino and Chopin(2009)]{RMC09}
Rue, H., Martino , S., and Chopin, N. (2009), ``Approximate Bayesian inference for latent Gaussian models by using integrated nested Laplace approximations," \textit{Journal of the Royal Statistical Society: Series B}, 71(2), 319--392.

\bibitem[Sato(2001)]{Sato01}
Sato, M. (2001), ``Online model selection based on the variational bayes,"  \textit{Neural Computation}, 13(7), 1649--1681.

\bibitem[Tokdar(2007)]{Tokdar07}
Tokdar, S. (2007), ``Towards a faster implementation of density estimation with logistic Gaussian process priors,"  \textit{Journal of Computational and Graphical Statistics}, 16, 633–-655.

\bibitem[Wang and Dunson(2011)]{WDU11}
Wang, L., and Dunson, D. B. (2011), ``Fast Bayesian Inference in Dirichlet Process Mixture Models," \textit{Journal of Computational and Graphical Statistics}, 20(1), 196--216. 

\bibitem[White and Ghosh(2009)]{WG09}
White, G., and Ghosh, S. K. (2009), ``A Stochastic Neighborhood Conditional Autoregressive Model for Spatial Data,"  \textit{Computational Statistics and Data Analysis}, 53(8), 3033--3046.

\end{thebibliography}
\end{document}